\documentclass[structabstract]{aa}
\usepackage{graphicx}
\usepackage{aalongtable}
\usepackage{lscape}
\usepackage{txfonts}
 \usepackage{subfig}
 \usepackage{natbib}
\bibpunct{(}{)}{;}{a}{}{,}

\begin{document}

%TITLE
\title{The TW Hydrae Association : trigonometric parallaxes and kinematic analysis 
\thanks{Based on observations performed at the European
Southern Observatory, Chile (79.C-0229, 81.C-0143, 82.C-0103, 83.C-0102, 84.C-0014).}}

%AUTHORS
\author{C. Ducourant \inst{1,2,3}
  \and R. Teixeira \inst{3,1}
  \and P. A. B. Galli \inst{3,1}
  \and J. F.  Le Campion\inst{1}
  \and A. Krone-Martins \inst{6,3,1}
  \and B. Zuckerman \inst{4}
  \and G. Chauvin \inst{5}
  \and I. Song \inst{7}
}

%INSTITUTES
\institute{
Univ. Bordeaux, LAB, UMR 5804, F-33270, Floirac, France.
\and
CNRS, LAB, UMR 5804, F-33270, Floirac, FranceObservatoire Aquitain des Sciences de l'Univers, CNRS-UMR 5804, BP 89, 33270 Floirac, France.
\and
Instituto de Astronomia, Geof\'isica e Ci\^encias Atmosf\'ericas, Universidade de S\~ao Paulo, Rua do Mat\~ao, 1226 - Cidade Universit\'aria, 05508-900 S\~ao Paulo - SP, Brazil.
\and
Department of Physics \& Astronomy, UCLA, Los Angeles, CA 90095 USA
\and
Laboratoire d'Astrophysique, Observatoire de Grenoble, 414, Rue de la piscine, 38400 Saint-Martin d'H\`eres, France
\and
SIM, Faculdade de Ci\^encias, Universidade de Lisboa, Ed. C8, Campo Grande, 1749-016, Lisboa, Portugal
\and
Department of Physics \& Astronomy, the University of Georgia, Athens, GA 30605 USA}

\date{Accepted Jan-7-2014}

%ABSTRACT
\abstract
%Context
{The nearby TW Hydrae Association (TWA) is currently a benchmark for the study of formation and evolution of young low-mass stars, circumstellar disks and the imaging detection of planetary companions. For such studies, it is crucial to evaluate the distance to group members in order to access their physical properties. Membership of several stars is strongly debated and age estimates vary from one author to  another with doubts about coevality.}
%Aims
{We revisit the kinematic properties of the TWA in light of new trigonometric parallaxes and proper motions to derive the dynamical age of the association and physical parameters of kinematic members.}
%Methods
{Using observations performed with the NTT/ESO telescope we measured trigonometric parallaxes and proper motions for 13 stars in TWA. }
%Results
{With the convergent point method we identify a co-moving group with 31 TWA stars. We deduce kinematic distances for 7 members of the moving group that lack trigonometric parallaxes. A traceback strategy is applied to the stellar space motions of a selection of 16 of the co-moving objects with accurate and reliable data yielding a dynamical age for the association of $t\simeq 7.5~\pm 0.7$~Myr. Using our new parallaxes and photometry available in the literature we derive stellar ages and masses  from theoretical evolutionary models.}
%Conclusions
{With new parallax and proper motion measurements from this work and current astrometric catalogs we provide an improved and accurate database for TWA stars to be used in kinematical analysis. We conclude that the dynamical age obtained via traceback strategy is  consistent with previous age estimates for the TWA, and is also compatible with the average ages derived in the present paper from evolutionary models for pre-main sequence stars.}

%KEYWORDS
\keywords{Astrometry : parallaxe and proper motion --  Stars: pre main sequence, brown dwarfs -- Galaxy: open cluster and association, TW Hydrae Association}

\maketitle

%----------------------------------------------------------------------------------------------------------------
%							INTRODUCTION
%----------------------------------------------------------------------------------------------------------------
\section{Introduction}

The discovery of nearby young stars, brown dwarfs and extrasolar planets has grown substantially in the last decade. In this context, the possibility of accurately determining their physical properties has attracted particular interest to the solar neighborhood. Since the discovery of the young \citep[$t\simeq8$~Myr,][]{Ramiro(2006)}, and nearby \citep[$d\simeq50$~pc,][]{Zuckerman(2004)}
 TW Hydrae association (TWA) by \citet{Kastner(1997)}, important
progress has been made in the identification of young stars near the
Sun, more than 200 of which have been
cataloged. 

TWA is among the closest of the very young associations and for this reason it has been a benchmark for the study of stellar and sub-stellar formation and early evolution.  The study of the TWA region by Webb et al (1999) demonstrated the power of the ROSAT All-Sky X-ray survey to reveal members not only of TWA but also of other, subsequently identified, youthful nearby associations.  A few years later \citet{Gizis(2002)} identified two free-floating brown dwarfs members of TWA, one of which, 2M1207, soon become famous as the host of the first imaged planet-mass secondary (2M1207b) outside of our solar system \citep{Chauvin(2004)}.   Many TWA members exhibit a signature of dusty disks \citep[e.g.][]{Riaz(2008),Riaz(2012),Looper(2010a),Matthews(2007),Schneider(2012a), Schneider(2012b)}. The age of TWA corresponds to the time-scale of the end of disk accretion and giant planet building processes. Recent work of \citet{Bergin(2013)} presents evidence that the disk around TW~Hydrae (TWA~1) is still capable of forming a planetary system.  The youth and proximity of TWA presents a particularly favorable situation for spatial resolution of disk structures of tens to hundreds of AU.

Dynamical measurements of low-mass binaries, for example TWA 22 \citep{Bonnefoy(2009)} and TWA 5 \citep{Neuhauser(2010)}, may provide quasi-unique opportunities for derivation of individual masses necessary for the calibration of theoretical evolutionary models.  Such evolutionary models can also be calibrated by use of kinematic traceback, a technique that was employed by \citep{Ortega(2002)} to derive an age of 12 Myr for the $\beta$ Pictoris group.  However, other, more recent traceback studies have produced older or indeterminate ages (see summary in Section 4 of \citet{Binks(2013)}. A major goal of the present paper is to improve the accuracy and reliability of the traceback age for the TWA.

Distance from Earth is a key parameter that enables the physical characterization of objects and kinematical studies of their origin. It is precisely the  high quality of astrometric measurements that makes the solar neighborhood, and consequently TWA, a precious laboratory. The association as a whole is also interesting since its kinematics and origin remain unclear. It is located far from molecular clouds and at the near boundary of the Lower Centaurus Crux (LCC) subgroup of the Scorpius-Centaurus (Sco-Cen) association.

Membership and age of TWA stars have been much debated \citep{Song(2003),Ortega(2004),Mamajek(2005),Ramiro(2006),Teixeira(2009),Schneider(2012b), Weinberger(2013)}.  The conclusions of these authors relied heavily on the type and accuracy of the data they used. A rich discussion of the constitution of TWA and a useful source of data utilized by many authors is presented in \citet{Mamajek(2005)}. In his paper, Mamajek  collected proper motions from different sources with inhomogeneous quality. The situation for radial velocities was equivalent and trigonometric parallaxes were available for only 5 Hipparcos stars. An important step in evaluation of the age of the association was pioneered by \citet{Makarov(2005)} and further developed by  \citet{Ramiro(2006)} who applied a traceback strategy to  the five Hipparcos TWA stars. They derived the epoch of minimum volume corresponding to the dynamical age of the association. Since then, the number of identified TWA members has increased as well as the quality and availability of data. 

In this context, we present here trigonometric parallax and proper motion measurements for 13 TWA stars performed with the NTT/ESO telescope located at La Silla (Chile). The recent increase of TWA stars  with measured trigonometric parallaxes \citep[this work and][]{Weinberger(2013)}, and the publication of the astrometric proper motion catalogues, SPM4 \citep{Girard(2011)} and UCAC4 \citep{Zacharias(2013)}, allows us to set up an extensive and accurate database for TWA stars. The PPMXL catalogue \citep{Roeser(2010)} was excluded from the present work since it provides proper motions of much lower internal accuracy than the two others catalogues. 

Using a convergent point analysis \citep{Galli(2012)} we identify a group of co-moving stars in TWA. Based on a traceback strategy we derive a core-group converging back in time towards a minimum volume in space that corresponds to the dynamical age of the association.

This paper is organized as follows. In Sect.~2 we present the observational material and describe the reduction procedure that leads to the astrometric and photometric properties of our targets. In Sect.~3 we present an updated astrometric database for the 34 proposed members of the association (TWA~1 - TWA~34). Sect.~4 describes our convergent point analysis and the determination of kinematic parallaxes for group members with unknown trigonometric parallax. We present in Sect.~5 a traceback analysis leading to a convincing traceback age for a core-group of the association.  Sect.~6 presents an HR-diagram of TWA along with age and mass estimates as derived from stellar evolutionary models. Our conclusions are summarized in Sect.~7.

%----------------------------------------------------------------------------------------------------------------
%								DATA	
%----------------------------------------------------------------------------------------------------------------
\section{Data}
For the present work we set up a list of all TWA members without parallax measurements.  Excluded were some resolved tight binaries for which an astrometric solution might be problematic and objects with (too) large photometric distances. TWA~1 and TWA~9A, which benefit from a Hipparcos parallax, were included in our final list as control stars. The list was then reduced to 15 stars during the observations to fit the allocated time. Two stars were observed but are not presented in this paper since a
reasonable solution could not be derived.  The remaining 13 objects are presented in Table  \ref{astrometry}.

\subsection{Observations}

Astrometric and photometric (V, R, I) observations were performed  in direct imaging mode with the ESO-NTT telescope. For the astrometric project, nine observational epochs were acquired with a total of 36 half-nights spread over almost three years between 2007 and 2010. A set of 3730 exposures were taken, concentrated in 13 directions corresponding to the selected members of TWA  for which a parallax measurement was required. Observational epochs were required at specific dates to maximize the parallactic factors of most targets. Given that the targets were spread over 3 hours in right ascension, we had to find a compromise between the maximization of parallax factors and the observability of all targets during a night. All observations were realized around transit to minimize the differential color refraction effects (DCR). Multiple exposures taken over three nights were performed at each epoch to average atmospheric effects and to enhance signal to noise ratio (S/N). For most objects, two exposure times were selected, a short one selected to optimize S/N of the bright targets and a longer one for an optimized S/N of the faint surrounding stars.

We devoted extreme efforts to observe during transit to minimize the zenith distance  and consequently DCR effects that could induce a factious parallax to a target  with a significantly different color than its surrounding background stars. 

The program started with the SUSI2 instrument (observations in 2007) but  unfortunately SUSI2 was decommissioned and the program was transported to the EFOSC2 instrument in 2008. The change of instrument had repercussion on the quality of the observations and the precision goal of the project was degraded by almost a factor 2. 

Frames were measured using the {\tt DAOPHOT II} package \citep{Stetson(1987)}, fitting a stellar point-spread function for each frame. Finally we created catalogs of measured positions $(x,y)$, internal magnitudes and associated errors for all objects in each frame.

%----------------------------------------------------------------------------------------------------------------
\subsection{Trigonometric parallax  and proper motion determination}

The catalogs that issued from the CCD frames are cross-correlated and compiled in a meta-catalog containing, for each object in a field, its measurements on each CCD-frame.

A frame, hearafter \textit{master frame}, is then selected among the various observations to compute equatorial coordinates of each object of the field, using 2MASS \citep{2MASS} as a reference catalog. These equatorial coordinates are necessary for the parallactic factors calculations and for this reason it is convenient to work in a frame oriented in equatorial coordinates. This step is equivalent to scale (size of the pixel) and rotate the master frame to align it on the axes of the 2MASS catalog. The selection of the master frame is a delicate problem since it will define, via the reference catalog used, the reference frame on which each CCD-frame will be later projected. It is then crucial that it is as free as possible of distortion and that it contains as many stars as possible, of various magnitudes so that the classical reduction performed with 2MASS be as accurate as possible. 

All measurements of a field are globally reduced through a block-adjustment type iterative procedure described in \citep{Ducourant(2007),Ducourant(2008)}. The philosophy of this treatment is to compute simultaneously the unknown parameters of all stars (correction to standard coordinates, proper motions, parallaxes) and the unknown plate parameters of all frames. The system is overdetermined and it would be in principle possible to derive by one inversion the whole of parameters. 

Nevertheless, the system of normal equations will be of $2N_{*}N_{F}$ equations for $5N_{*}+6N_{F}$ unknowns if we consider $N_{F}$ frames containing $N_{*}$ stars, leading to a matrix of large dimensions (typically few tens of thousands equations for thousands of unknowns for a few hundreds frames containing 50 to 100 stars). The size of such a large matrix leads the user to prefer an iterative approach to solve the system. We use a Gauss-Seidel method allowing one to first determine the plate constants assuming an a-priori value for all stellar parameters. The stellar parameters are determined for each star using modified plate coordinates, the plate constants are re-computed, and so on until convergence which is generally reached after 2 or 3 iterations. 

This system is ill conditioned which means that  the iterative solving of the normal equations will converge towards one of the various solutions of the problem but not necessarily to the most physical one. To constrain the convergence to the ``most physical'' solution, one must add a constraint. Generally one assumes that the sum of parallaxes in the field should be zero (excluding the target from this mean); the same condition applies for proper motions. Convergence is reached after a few iterations and a relative solution (proper motion and parallax) is derived for all stars in the field.

A statistical conversion from relative to absolute parallax and proper motions --
based on the Besan\c con Galaxy model \citep{Robin(2003),Robin(2004)} -- is applied to derive final estimates of absolute proper motions and trigonometric parallaxes \citep[see][for a detailed description]{Ducourant(2007)}. 

The procedure described in this section is applied to all 13 TWA stars observed in our campaign. The results of this investigation, including the corrections applied to the relative quantities in order to derive absolute astrometric parameters, are given in Table~\ref{astrometry}.  

%TABLE 1
%ces rŽsultats proviennent de cross/article/TWApar.dat + twapar.prg
\begin{table*}[!ht]
\caption{\label{astrometry}Absolute trigonometric parallaxes and proper motions derived in this work for the 13 TWA stars. We provide for each star its position (epoch=2000.0), absolute parallax, distance, absolute proper motions and the $\Delta$ corrections applied to relative values of parallax and proper motions. For example, to obtain absolute parallax for TWA 1, $\pi_{relative}$ + 0.78 = $\pi$ given in column 4. }
\begin{tabular}{lccccccccc}
\hline
Star&$\alpha$&$\delta$&$\pi$&$d$&$\mu_{\alpha}\cos\delta$&$\mu_{\delta}$&$\Delta\pi$&$\Delta\mu_{\alpha}\cos\delta$&$\Delta\mu_{\delta}$\\
&(h:m:s)&($^{\circ}$ $^\prime$ $^\prime$$^\prime$)&(mas)&(pc)&(mas/yr)&(mas/yr)&(mas)&(mas/yr)&(mas/yr)\\
\hline
 TWA 1  &11 05 51.97  &-34 42 16.9  &$20.8\pm4.0$ & $48.1\pm 9.3$ & $-68.4\pm1.5$ & $-8.0\pm 1.5 $&0.78 &-7.18 & 1.11 \\ 
 TWA 2  &11 09 13.88  &-30 01 39.7  &$24.0\pm4.8$ & $41.6 \pm8.3$ & $-87.7\pm2.3$ & $-7.9\pm 2.3$ &0.81 &-7.20 & 0.24 \\ 
 TWA 5   &11 31 55.46  &-34 36 28.8  &$20.5\pm2.4$ & $48.7 \pm5.7$ & $-75.8\pm1.0$ &$-18.3\pm 1.0$ &0.73 &-7.00 & 0.28 \\ 
 TWA 7  &10 42 30.20  &-33 40 16.0  &$29.0\pm2.1$ & $34.5 \pm2.5$ &$-114.4\pm0.8$ &$-19.1\pm 0.8$ &0.85 &-7.24 & 1.45 \\ 
 TWA 8A  &11 32 41.32  &-26 51 55.6  &$23.4\pm2.0$ & $42.8\pm3.7$ & $-87.1\pm0.8$ &$-28.0\pm 0.8$ &0.89 &-7.62 &-1.00 \\ 
 TWA 8B  &11 32 41.23  &-26 52 08.7  &$25.9\pm2.0$& $38.6\pm3.0$ & $-86.5\pm0.9$ &$-25.0\pm 0.9$ &0.89 &-7.62 &-1.00 \\ 
 TWA 9A  &11 48 24.22  &-37 28 49.2  &$19.1\pm2.7$ & $52.3\pm7.4$ & $-53.1\pm3.9$ &$-24.9\pm 3.9$ &0.71 &-7.05 & 0.10 \\ 
 TWA 9B  &11 48 23.77  &-37 28 48.3  &$19.2\pm1.1$& $52.1\pm3.0$ & $-51.0\pm0.6$ &$-18.1\pm0.6$ &0.71 &-7.05 & 0.10 \\
 TWA 10  &12 35 04.31  &-41 36 38.3  &$16.2 \pm1.0$ & $61.5 \pm3.8$ &$ -64.6\pm0.4$ &$-30.3\pm 0.4$ &0.55 &-6.58 &-0.52 \\ 
 TWA 12  &11 21 05.55  &-38 45 16.2  &$15.4\pm1.7$ & $65.1\pm7.2$ &$ -66.2\pm0.5$ &$-7.4\pm 0.5$ &0.67 &-6.80 & 0.70 \\ 
 TWA 21  &10 13 14.85  &-52 30 54.1  &$19.8 \pm1.4$ & $50.4 \pm3.6$ &$ -61.3\pm0.6$ &$10.6\pm 0.6$ &1.03 &-7.49 & 3.39 \\
 TWA 23  &12 07 27.44  &-32 47 00.0  &$20.6 \pm1.8$ & $48.4\pm4.2$ & $-75.8\pm0.9$ &$-25.7\pm 0.9$ &0.74 &-7.23 &-0.79 \\ 
 TWA 26  &11 39 51.21  &-31 59 21.2  &$26.2 \pm1.1$ & $38.1\pm1.6$& $-93.3 \pm0.5$ &$-27.5\pm 0.5$ &0.64 &-6.30 & 0.11 \\ 
     \hline
  \end{tabular}
\end{table*}

%----------------------------------------------------------------------------------------------------------------
\subsection{Astrometric validation}

In the following we compare the parallaxes and proper motions derived in this work with published results in order to evaluate our external errors. 

\begin{figure}[!htp]
\begin{center}
\includegraphics[width=0.49\textwidth]{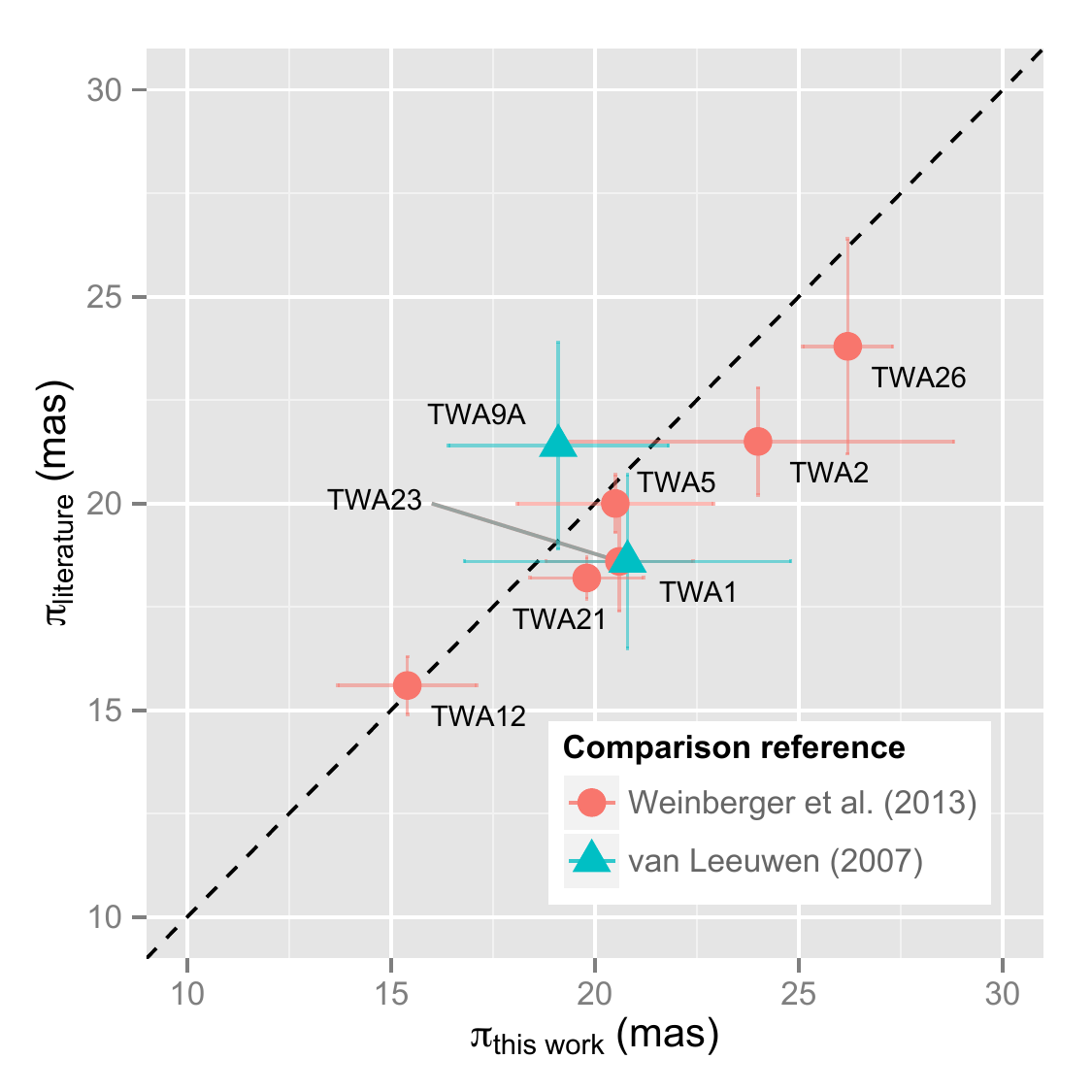}
\caption{Comparison of trigonometric parallaxes determined in this work  with the published parallaxes given in \textit{Hipparcos} \citep{HIP07} and \citet{Weinberger(2013)}. The dotted line represents perfect correlation. TWA 23, which was measured by us and by Weinberger et al, sits essentially under TWA 1.  The error bar on our measurement is +/-1.8 mas. 
\label{comparpi}}
\end{center}
\end{figure}

In the case of  trigonometric parallaxes, a comparison with \textit{Hipparcos} is difficult, because only five TWA stars are in that catalog. Recently, in a project parallel to the one presented here, \citet{Weinberger(2013)} radically improved the situation by publishing parallaxes for 14 TWA systems (16 stars).  A comparison of Table~\ref{astrometry} with the re-reduction of \textit{Hipparcos} \citep{HIP07}  is restricted to only two stars (TWA~1 and TWA~9A), while there are six stars in common with \citet{Weinberger(2013)}.  For five of these six the internal errors listed in our Table~\ref{astrometry} are larger than the internal errors listed in their Table 2.  Notwithstanding that all six parallaxes appear to be consistent at the one-sigma level (Fig.~\ref{comparpi}), when comparing our results with \citet{Weinberger(2013)} we notice an apparent systematic offset of unknown origin. The size of the offset, a few mas, is comparable to the mean internal precision (2.2 mas) of the trigonometric parallaxes derived in the present paper.

%FIGURE 1

\begin{figure*}[!]
\includegraphics[width=0.49\textwidth]{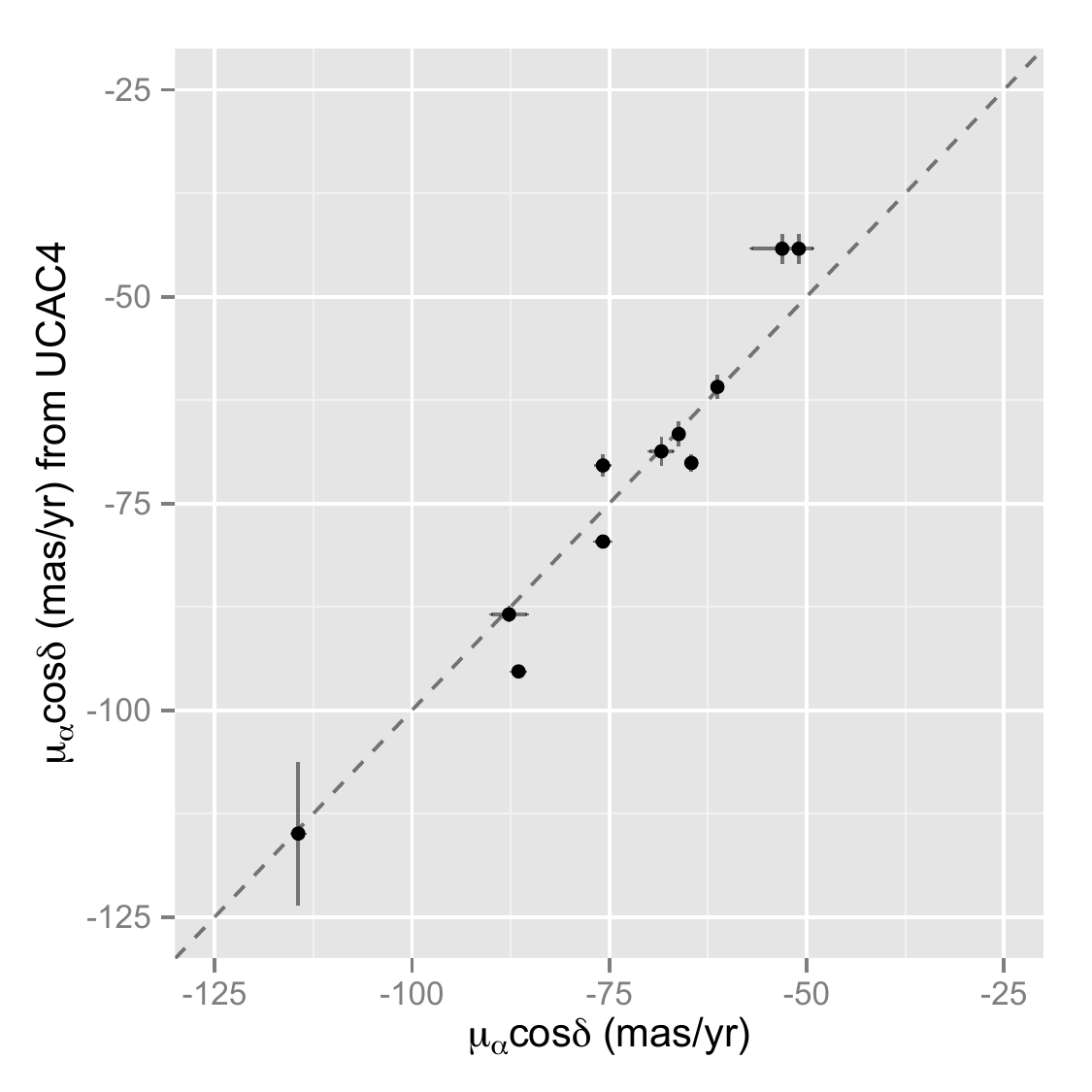}
\includegraphics[width=0.49\textwidth]{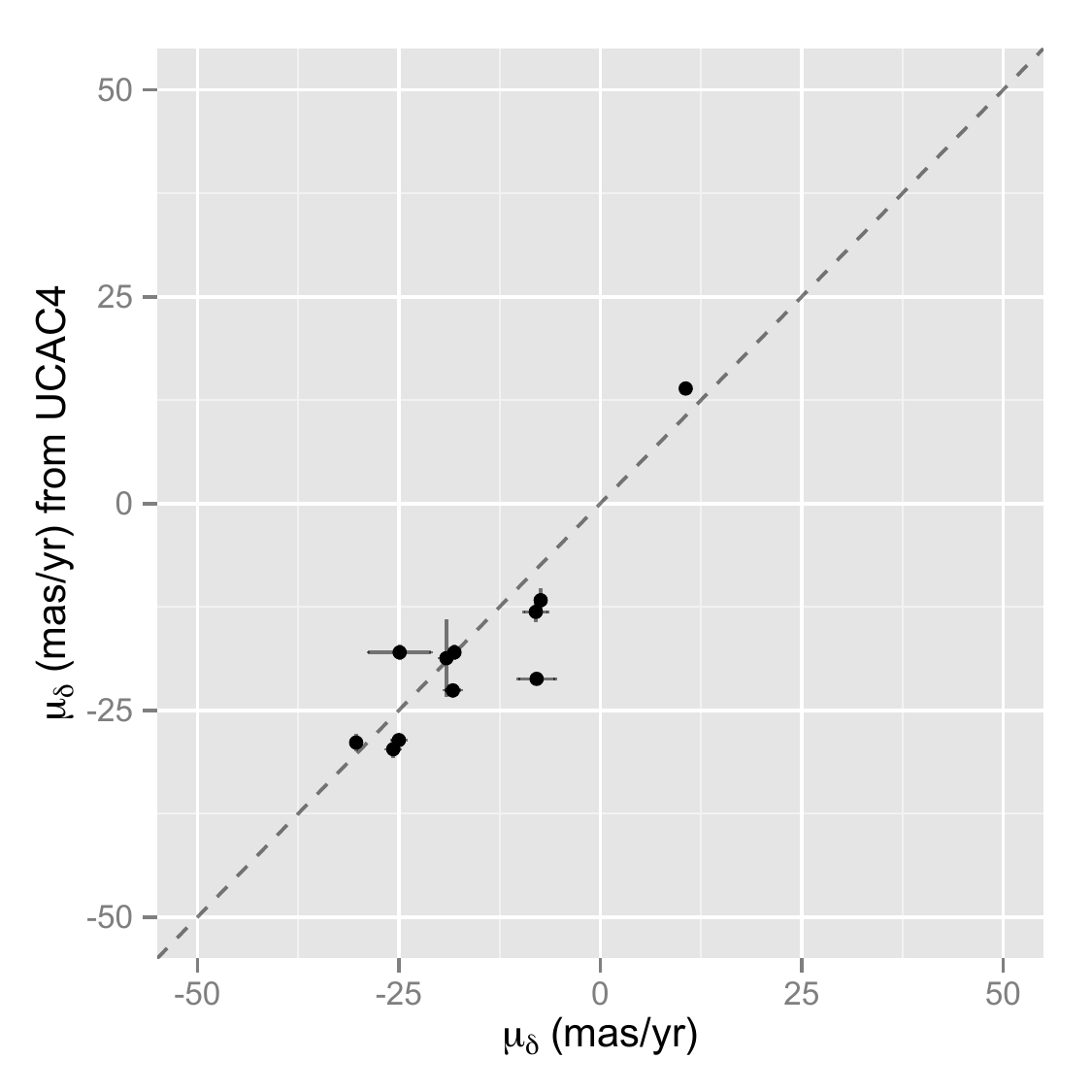}
\caption{\label{comparmu} Comparison of the proper motions derived in our work with those from UCAC4. The dotted line indicates the expected results for perfect correlation. In the left panel, abscissa entries are right ascension values from column 6 in Table~\ref{astrometry}, and in the right panel declination values from column 7.}
 \end{figure*}

We compared our proper motions with those from the recent release of the UCAC4 catalog that provides a dense and precise astrometric source of data with an announced precision of 1-10~mas/yr depending on magnitude and observing history. The mean internal precision of the proper motions derived in the present paper is 1.0~mas/yr. Fig.~\ref{comparmu} shows the comparison of the measured proper motions given in Table~\ref{astrometry} with UCAC4 proper motions. We observe a reasonable agreement in both coordinates and no systematic trend. However large discrepancies (beyond three sigmas) can be observed. The origin of such discrepancies is  unclear and can be multiple. Binarity is probably one cause. It is also probable that the formal errors of both works are occasionally underestimated. Nevertheless one must keep in mind that in the present work, the timebase of observations is three years, optimized for parallax work but which is rather short for accurate proper motion determination, especially for multiple systems. For this reason,  in most cases we used in our kinematic analysis proper motions issued from UCAC4. 

%FIGURE 2
 
%----------------------------------------------------------------------------------------------------------------
\subsection{Photometry}

Photometric data were acquired during three consecutive nights (3, 4, 5 April 2009) using the Bessel V, R and Gunn~i ESO filters (ESO\#641,642,705). We present in Table~\ref{photometry} the photometry derived for the TWA stars observed in this work . 

TWA~7 has a V magnitude that is not in accordance with Tycho-2 \citep{Hog(2000)}, but in perfect agreement with the accurate photometric variability project of \citet{Messina(2010)}. It is probable that the faint object close to TWA~7 ($\delta\sim 2.4\arcsec$, \citet{Messina(2010)}) was  included in the Tycho-2 photometry. Our V and I magnitudes for TWA~8B do not agree with those published by \citet{Messina(2010)}. Exposure times in our work were optimized for the primary TWA~8A which is much brighter; this may explain the poor results obtained here for the secondary component. For TWA~9A and TWA~9B no reliable photometry could be obtained here. 

%TABLE 2
\begin{table*}[!]
\begin{center}
\caption{\label{photometry}
Photometry of the TWA stars observed in this work (except TWA~1). We provide for each star the V, R, I magnitudes derived in this paper, 2MASS JHK photometry \citep{2MASS} and comments on the multiplicity (B for binary and T for triple) of targets as summarized in  Appendix A of \citet{Messina(2010)}. The symbol ``*'' denotes rough photometry.
\vspace{0.5cm}}

\begin{tabular}{lcccccccccc}
\hline
Star&$V$&$R$&$Gunn~i$&$J$ (2MASS)&$H$ (2MASS)&$K$ (2MASS)&Multiplicity\\
&(mag)&(mag)&(mag)&(mag)&(mag)&(mag)\\
\hline
TWA 2    &  11.10  $\pm$ 0.03  & 10.04 $\pm$ 0.04 &    8.91 $\pm$ 0.05 &    7.629 &   6.927 &   6.710 & B\\
TWA 5   &  11.50   $\pm$ 0.03  & 10.40 $\pm$ 0.04 &    9.19 $\pm$ 0.05 &    7.669 &   6.987 &   6.745 & T\\
TWA 7    &  11.78  $\pm$ 0.03  & 10.61 $\pm$ 0.03 &    9.43 $\pm$ 0.04 &    7.792 &   7.125 &   6.899 & \\
TWA 8A  &  12.33  $\pm$ 0.03  & 11.18 $\pm$ 0.04 &    9.88 $\pm$ 0.05 &    8.337 &  7.663  &  7.430 & B\\
TWA 8B  &  18.8* & 16.6* &  15.3*  &    9.837 &   9.276 &   9.012 & B \\
TWA 10    &  12.91 $\pm$ 0.03  & 11.74 $\pm$ 0.03 &  10.45 $\pm$ 0.04 &    9.122 &   8.477 &   8.186 & \\
TWA 12    &  12.94 $\pm$ 0.07  & 11.81 $\pm$ 0.06 &  10.58 $\pm$ 0.06 &    8.999 &   8.334 &   8.053 & \\
TWA 21    &    9.85 $\pm$ 0.07  & 9.29   $\pm$ 0.06 &    8.75 $\pm$ 0.06 &    7.870 &   7.353 &   7.194 & \\
TWA 23    &  12.69 $\pm$ 0.03  & 11.45 $\pm$ 0.03 &  10.07 $\pm$ 0.04 &    8.618 &   8.025 &   7.751 &\\
TWA 26    &  17.25 $\pm$ 0.02  & 16.84 $\pm$ 0.03 &  16.46 $\pm$ 0.05 &  12.686 & 11.996 & 11.503 & \\
\hline
  \end{tabular}
\end{center}
\end{table*}

%----------------------------------------------------------------------------------------------------------------
%					AN UPDATED DATABASE FOR TWA	
%----------------------------------------------------------------------------------------------------------------
\section{An updated astrometric database for TWA}

Using the parallaxes and proper motions derived in this work, the recent results of \citet{Weinberger(2013)}, and the new release of SPM4 and UCAC4, we set up an updated astrometric database for TWA stars that allows us to revisit completely the kinematics of the association. The selection of data used in such analysis is a very important step, because published values can vary widely from one source to another. In the following we discuss our criteria for choosing (among various sources) parallaxes, proper motions and radial velocities for TWA stars. 

We use the trigonometric parallaxes from the new reduction of \textit{Hipparcos} \citep{HIP07} when available and combine them with the associated Tycho-2 proper motions. For the remaining TWA stars we use the trigonometric parallaxes derived in this work which we supplement with those provided by \citet{Weinberger(2013)}. The average parallax error is 1.5~mas (Table \ref{data}) yielding an average relative error of 8\%. 

In the case of proper motions, we searched the SPM4 and UCAC4 catalogs. For common stars in both catalogs we favored the one with the lowest errors on proper motions which in general corresponds to UCAC4. In the case of discrepant values we compared the proper motion values with other sources \citep[e.g., this work and][]{Weinberger(2013)} to decide between the two catalogs. For the specific cases of multiple systems that exhibit a poor or inconsistent solution (e.g., TWA~13 and TWA~15), and stars lacking information in global catalogs we adopted the proper motions from small field astrometry \citep[this work and][]{Weinberger(2013)} when no other reliable data were available. Doing so, the average error on proper motions is 1.8~mas/yr and 1.6~mas/yr, respectively, in right ascension and declination. This yields an average relative error on proper motions better than 4\%.  

A proper motion for TWA 29 of $(-89.4,-20.9) \pm(10, 10)$~mas/yr , was estimated by \citep{Schneider(2012a)} based on 2MASS and WISE positions.
However, the proper motion derived by \citet{Looper(2007)} from UKST observations combined with 2MASS and DENIS positions is radically different $(-220,+20) \pm (70, 100)$~mas/yr. So we decided to remeasure this proper motion using the ESO archive database (15 NTT/SOFI NIR images taken from 2003 to 2006) and the WISE ($\sim$2010), 2MASS($\sim$2000), DENIS($\sim$1999) and GSC2($\sim$1984) published positions. We derived a proper motion of $(-71,-23) \pm(7, 3)$~mas/yr. This object has been included in our analysis despite the large error-bars. 

For radial velocities we collected the available measurements in the literature and we rarely had to perform a selection between the various sources, because they exist in small number as compared to sources for proper motions. In the case of  TWA~4, a tight quadruple system \citep{Messina(2010)}, values found in the literature refer either to the A or B component while here we are using a Tycho2 proper motion for this star that refers to the photo-center of the system. We therefore adopted the radial velocity of the center of mass of the system provided by \citet{Torres(2003)}. 
We present the results of our data compilation for TWA stars in Table~\ref{data}. In the following sections we investigate the kinematics of the association and the properties of individual stars using data presented in this table. 

%TABLE 3
\begin{table*}[!htp]
\begin{center}
\caption{\label{data}
Selected astrometric data for the 34~TWA stars (and stellar systems). }
\resizebox{19cm}{!}{
\begin{tabular}{lccccccccccccll}
\hline
(1) & (2) & (3) &  (4) &  (5) &  (6) &  (7) &  (8) &  (9) &  (10) &  (11) &  (12) &  (13) & (14) \\
Star&$\alpha$&$\delta$&$\sigma_{\alpha}$&$\sigma_{\delta}$&$\mu_{\alpha}\cos\delta$&$\mu_{\delta}$&Ref.&$V_{r}$&Ref.&$\pi$&Ref.& CP& Traceback \\
&(h:m:s)&($^{\circ}$ $^\prime$ $^\prime$$^\prime$)&(mas)&(mas)&(mas/yr)&(mas/yr)&&(km/s)&&(mas)&&Member.& Age deter.\\
\hline
TWA~1    &11 01 51.905& -34 42 17.03&  05& 04&  -73.4 $\pm$ 2.3 & -17.5 $\pm$ 2.3 & 1 & 12.66 $\pm$ 0.22& 7 & 18.6 $\pm$  2.1 & 12 & Y & + \\
TWA~2    &11 09 13.798& -30 01 39.88&  16& 13&  -88.4 $\pm$ 1.0 & -21.2 $\pm$ 0.8 & 2 & 10.58 $\pm$ 0.51& 8 & 24.0 $\pm$  4.8 & 3 & Y & + \\
TWA~3    &11 10 27.894& -37 31 51.97&  14& 15& -105.9 $\pm$ 0.9 & -17.3 $\pm$ 1.0 & 2 &  9.52 $\pm$ 0.86& 8 &28.3 $\pm$  1.6* & 3 & Y & + \\
TWA~4    &11 22 05.288& -24 46 39.74&  04& 18&  -91.7 $\pm$ 1.6 & -31.1 $\pm$ 1.4 & 1 & 9.20 $\pm$ 0.10& 7 & 22.3 $\pm$  2.3   & 12 & Y & + \\
TWA~5    &11 31 55.260& -34 36 27.25&  10& 12&  -79.6 $\pm$ 0.8 & -22.6 $\pm$ 0.9 & 2 & 13.30 $\pm$ 2.00& 8 & 20.5 $\pm$  2.4 & 3 & Y & r \\
TWA~6    &10 18 28.701& -31 50 02.86&  14& 12&  -55.0 $\pm$ 1.2 & -19.8 $\pm$ 1.0 & 2 & 16.90 $\pm$ 5.00&15 & 14.9 $\pm$ 4.4*& 3 &Y  &r    \\
TWA~7    &10 42 30.100& -33 40 16.28& 128& 63& -114.4 $\pm$ 0.8 & -19.1 $\pm$ 0.8 & 3 & 12.21 $\pm$ 0.24& 8 & 29.0 $\pm$2.1 & 3 & Y & + \\
TWA~8A   &11 32 41.263& -26 51 55.99&  09& 09&  -90.6 $\pm$ 0.9 & -32.1 $\pm$ 0.8 & 4 &  8.34 $\pm$ 0.48& 8 & 23.4 $\pm$  2.0 & 3 & Y & + \\
TWA~8B   &11 32 41.172& -26 52 09.13&  33& 78&  -93.2 $\pm$ 1.3 & -27.5 $\pm$ 1.2 & 4 &  8.93 $\pm$ 0.27& 8 & 25.9 $\pm$  2.0 & 3 & Y & + \\
TWA~9A   &11 48 24.229& -37 28 49.11&  06& 03&  -55.4 $\pm$ 2.3 & -17.7 $\pm$ 2.3 & 1 &  9.46 $\pm$ 0.38& 7 & 21.4 $\pm$  2.5 & 12 & Y &r  \\
TWA~9B   &11 48 23.732& -37 28 48.50&  27& 27&  -51.0 $\pm$ 0.6 & -18.1 $\pm$ 0.6 & 3 & 11.30 $\pm$ 2.00&15 & 19.2 $\pm$  1.1 & 3 & Y &r   \\
TWA~10   &12 35 04.254& -41 36 38.64&  16& 14&  -64.6 $\pm$ 0.4 & -30.3 $\pm$ 0.4 & 3 &  6.75 $\pm$ 0.40& 8 & 16.2 $\pm$  1.0 & 3 &Y? &--   \\
TWA~11A  &12 36 01.031& -39 52 10.23&  02& 02&  -53.3 $\pm$ 1.3 & -21.2 $\pm$ 1.1 & 1 &  9.40 $\pm$ 2.30& 7 & 13.7 $\pm$  0.3 & 12 & Y & + \\
TWA~11B  &12 36 01.031& -39 52 10.23&  02& 02&  .......... & .......... & - &  9.00 $\pm$ 1.00& 7 & ..........& - &Y  &--     \\
TWA~11C  &12 35 48.939& -39 50 24.50&  26& 26&  -48.6 $\pm$ 1.7 & -21.3 $\pm$ 1.6 & 2 & ..........&-& 14.5 $\pm$  0.5 & 5 &Y &-- \\
TWA~12   &11 21 05.484& -38 45 16.51&  16& 17&  -66.6 $\pm$ 1.5 & -11.7 $\pm$ 1.5 & 2 & 13.12 $\pm$ 1.59& 8 & 15.4 $\pm$  1.7 & 3 & Y & + \\
TWA~13A  &11 21 17.219& -34 46 45.47&  13& 16&  -66.4 $\pm$ 2.4 & -12.5 $\pm$ 1.8 & 5 & 11.67 $\pm$ 0.64& 7 & 18.0 $\pm$  0.7 & 5 & Y & + \\
TWA~13B  &11 21 17.446& -34 46 49.83&  13& 16&  -68.0 $\pm$ 3.1 & -11.0 $\pm$ 2.7 & 5 & 12.57 $\pm$ 0.50& 7 & 16.8 $\pm$  0.7 & 5 & Y & + \\
TWA~14   &11 13 26.221& -45 23 42.75&  14& 14&  -43.9 $\pm$ 1.4 &  -7.4 $\pm$ 1.4 & 2 & 15.83 $\pm$ 2.00& 8 & 10.4 $\pm$  1.2 & 5 & Y & + \\
TWA~15A  &12 34 20.649& -48 15 13.48&  17& 17&  -37.5 $\pm$ 2.4 & -10.4 $\pm$ 2.0 & 5 & 11.20 $\pm$ 2.00&15 &  9.1 $\pm$  1.7 & 5 & Y & + \\
TWA~15B  &12 34 20.473& -48 15 19.59&  22& 17&  -36.5 $\pm$ 2.9 &  -9.9 $\pm$ 2.8 & 5 & 10.03 $\pm$ 1.66& 8 &  8.6 $\pm$  1.6 & 5 & Y & +\\
TWA~16   &12 34 56.303& -45 38 07.63&  18& 11&  -47.5 $\pm$ 1.3 & -20.2 $\pm$ 0.8 & 2 &  9.01 $\pm$ 0.42& 8 & 12.8 $\pm$  0.5 & 5 &Y?  &--  \\
TWA~17   &13 20 45.388& -46 11 37.72&  16& 10&  -31.3 $\pm$ 1.1 & -17.7 $\pm$ 1.0 & 2 &  4.60 $\pm$ 6.00 &15 & ..........& - &N &--	  \\
TWA~18   &13 21 37.225& -44 21 51.84&  14& 14&  -32.1 $\pm$ 1.1 & -20.4 $\pm$ 1.1 & 2 &  6.90 $\pm$  3.00&15 & ..........& - &N &--	  \\
TWA~19A  &11 47 24.545& -49 53 03.01&  03& 03&  -33.7 $\pm$ 1.1 &  -9.1 $\pm$ 1.1 & 1 & 11.50 $\pm$ 3.80& 7 & 10.9 $\pm$  1.3 & 12 &N &--  \\
TWA~19B  &11 47 20.642& -49 53 04.31&  44& 47&  -21.6 $\pm$ 1.7 & -23.4 $\pm$ 1.6 & 4 & 15.20 $\pm$ 2.00&15 & ..........& - &N &--  \\
TWA~20   &12 31 38.068& -45 58 59.47&  12& 12&  -63.5 $\pm$ 1.1 & -27.8 $\pm$ 1.1 & 2 &  8.10 $\pm$ 4.00&15 & 12.9 $\pm$  0.6 & 5 &Y? &--    \\
TWA~21   &10 13 14.774& -52 30 53.95&  16& 10&  -60.9 $\pm$ 1.5 &  13.9 $\pm$ 0.8 & 2 & 17.50 $\pm$ 0.80&10 & 19.8 $\pm$  1.4 & 3 & Y &r \\
TWA~22   &10 17 26.905& -53 54 26.42&  18& 18& -175.8 $\pm$ 0.8 & -21.3 $\pm$ 0.8 &13 & 13.57 $\pm$ 0.26& 8 & 57.0 $\pm$  0.7 & 13 & N &--  \\
TWA~23   &12 07 27.377& -32 47 00.25&  18& 15&  -75.8 $\pm$ 0.9 & -25.7 $\pm$ 0.9 & 3 &  8.52 $\pm$ 1.20& 8 & 20.6 $\pm$  1.8 & 3 & Y & + \\
TWA~24   &12 09 41.861& -58 54 45.08&  12& 18&  -35.2 $\pm$ 1.8 & -14.9 $\pm$ 2.1 & 2 & 11.90 $\pm$ 0.90&10 & ..........& - &N &-- 	  \\
TWA~25   &12 15 30.723& -39 48 42.59&  12& 13&  -73.2 $\pm$ 0.8 & -27.7 $\pm$ 0.8 & 2 &  9.20 $\pm$ 2.10&10 & 18.5 $\pm$  1.2 & 5 &Y? &--     \\
TWA~26   &11 39 51.140& -31 59 21.50&  60& 60&  -93.3 $\pm$ 0.5 & -27.5 $\pm$ 0.5 & 3 & 11.60 $\pm$ 2.00&14 & 26.2 $\pm$  1.1 & 3 &Y &r\\
TWA~27   &12 07 33.467& -39 32 54.00&  60& 60&  -64.2 $\pm$ 0.4 & -22.6 $\pm$ 0.4 & 3 & 11.02 $\pm$ 2.00&14 & 19.1 $\pm$  0.4 & 3 &Y &r   \\
TWA~28   &11 02 09.833& -34 30 35.53&  60& 60&  -67.2 $\pm$ 0.6 & -14.0 $\pm$ 0.6 &13 & ..........& -& 18.1 $\pm$  0.5 & 13 & Y &--\\
TWA~29   &12 45 13.798& -44 28 47.63&  44& 38&  -71.0 $\pm$ 7.0 & -23.0 $\pm$ 3.0 & 3  & ..........&- &  12.7 $\pm$ 2.1     & 5      & Y &--\\ 
TWA~30A  &11 32 18.315& -30 19 51.85&  20& 20&  -87.8 $\pm$ 1.3 & -25.2 $\pm$ 1.3 & 2 & 12.30 $\pm$ 1.50&11 &18.0 $\pm$  2.2* & 3 & Y &r\\
TWA~30B  &11 32 16.921& -30 18 10.53&  21& 21&  -83.0 $\pm$ 9.0 & -30.0 $\pm$ 9.0 &11 & 12.00 $\pm$ 3.00&11 &17.8 $\pm$  4.8* & 3 & Y &r\\
TWA~31   &12 07 16.547& -32 30 22.36&  10& 10&  -42.0 $\pm$ 6.0 & -36.0 $\pm$ 3.0 & 6 & 10.47 $\pm$ 0.41& 8 & ..........&  - &N &-- 	  \\
TWA~32   &12 26 51.367& -33 16 12.54&  32& 79&  -59.7 $\pm$ 2.4 & -22.9 $\pm$ 2.2 & 4 &  7.15 $\pm$ 0.26& 8 &13.0 $\pm$  0.7* & 3 & Y & +\\
TWA~33&11 39 33.846&-30 40 00.34&50&45&-73.3 $\pm$ 2.9&-25.4 $\pm$ 2.6&4&..........&-&19.0 $\pm$ 1.2*&3&Y &-- \\
TWA~34&10 28 45.790&-28 30 37.53&38& 35&-68.6 $\pm$ 2.7&-11.4 $\pm$ 2.5&4&..........&-&20.0 $\pm$ 1.3*&3&Y &--\\

\hline
\end{tabular}
  }
\end{center}

\tablefoot{We provide for each star  in col. 2, 3, 4 and 5 its position (epoch=2000.0) from UCAC4 whenever available otherwise from SPM4 or 2MASS. The proper motion, radial velocity, parallax and the corresponding references are given in col. 6, 7, 8, 9, 10, 11, 12. Parallaxes with a ``*'' symbol are kinematic parallaxes derived in Sect.~4.3. Column 13 gives the membership status derived from the CP search method (`Y' = member, `N' = non-member, `Y?' = possible member) as defined in Sect.~4.4. Column 14 summarizes the participation of CP members to the trace-back age determination (see Section 5.2 for details) (+=participation, r=rejected, --:not member or lack of data).
}

\tablebib{
(1)~Tycho2 \citep{Hog(2000)};
(2)~UCAC4 \citep{Zacharias(2013)};
(3)~This paper or \citet{Ducourant(2008)};
(4)~SPM4 \citep{Girard(2011)};
(5)~\citet{Weinberger(2013)};
(6)~\citet{Schneider(2012a)};
(7)~\citet{Torres(2003)};
(8)~\citet{Shkolnik(2011)};
(9)~\citet{Torres(2008)};
(10)~\citet{Song(2003)}; 
(11)~\citet{Looper(2010a),Looper(2010b)};
(12)~\textit{Hipparcos} \citep{HIP07};
(13)~\citet{Teixeira(2008),Teixeira(2009)};
(14)~\citet{Mohanty(2003)};
(15)~\citet{Reid(2003)}.

}
\end{table*}

%----------------------------------------------------------------------------------------------------------------
%			CONVERGENT POINT AND MEMBERSHIP ANALYSIS	
%----------------------------------------------------------------------------------------------------------------
\section{Convergent point and membership analysis}

In the following we use the convergent point (CP) search method to identify a reliable moving group in the sample of stars listed in Table~3 that will be the starting point of our traceback analysis (see Sect.~6) to determine the dynamical age of the association. 

\subsection{Convergent point\label{cp}}

To accurately determine the CP position and perform a membership analysis of TWA stars we apply our new CP search method  \citep{Galli(2012)} to the proper motion data given in Table \ref{data}. The method takes the velocity dispersion and mean distance of the moving group as input parameters. The intrinsic velocity dispersion of TWA is expected to be $\sigma_{v}\leq 1$~km/s  and the mean distance is $d\simeq50$~pc \citep[see][]{Mamajek(2005)}. While the velocity dispersion term in the CP analysis allows us to identify those group members that do not show strict convergence to the CP, it also drives the method to include some additional stars that do not belong to the moving group \citep{deBruijne(1999a), Galli(2012)}. Thus, to define a secure group of comoving stars we consider in a first step $\sigma_{v}=0$~km/s and run the CP search method on the sample of classical TWA stars (TWA~1 - TWA~34). The results of membership analysis can also be understood in terms of the stop parameter $\epsilon_{min}$ \citep[see][for more details]{Galli(2012)} that should allow us to find the largest number of moving group members with the least contamination by field stars. To better compare our results with the previous CP analysis of TWA stars performed by \citet{Mamajek(2005)} we adopt a rejection threshold of 5\% following his procedure. Doing so, we find a moving group of 18~stars (TWA~1, TWA~3, TWA~4, TWA~7, TWA~12, TWA~13A, TWA~13B, TWA~15A, TWA~15B, TWA~21, TWA~23, TWA~26, TWA~29, TWA~30A, TWA~30B, TWA~32, TWA~33 and TWA~34)
that shows strict convergence and yields the best CP estimate to date for TWA located at
\begin{center}
$(\alpha_{cp},\delta_{cp})=(102.4^{\circ},-27.3^{\circ})\pm(1.4^{\circ},0.6^{\circ})$,
\end{center}
with chi-squared statistics $\chi^{2}_{red}=1.1$ (i.e., $\chi^{2}/\nu=17.4/16$).

We estimate the velocity dispersion of the moving group using Eq.~(19) of \citet{deBruijne(1999b)} that is given by
\begin{equation}\label{eq.1}
\overline{\mu_{\perp}^{2}}=\overline{(A^{-1}\, \pi\, \sigma_{v})^{2}}+\overline{\sigma_{\mu_{\perp}}^{2}}\,,
\end{equation}
where A = 4.74047~km~yr/s is the ratio of one astronomical unit in km to the number of
seconds in one Julian year. We compute $\mu_{\perp}$, the stellar proper motion component directed perpendicular to the great circle that joins the star and the CP, using the following transformation 
\begin{equation}
\begin{array}{c}
\left(
  \begin{array}{r}
    \mu_{\parallel} \\
    \mu_{\perp} \\
  \end{array}
\right)
=
\left(
  \begin{array}{cc}
    \sin\theta & \cos\theta \\
    -\cos\theta & \sin\theta \\
  \end{array}
\right)
\left(
  \begin{array}{c}
    \mu_{\alpha}\cos\delta \\
    \mu_{\delta} \\
  \end{array}
\right),
\end{array}
\end{equation}\newline
where $\theta$ is given by  \citep[see also][]{Galli(2012),deBruijne(1999a)}
\begin{equation}\label{eq.2.2}
 \tan\theta=\frac{\sin(\alpha_{cp}-\alpha)}{\cos\delta\tan\delta_{cp}-\sin\delta\cos(\alpha_{cp}-\alpha)}{
}\, .
\end{equation}\newline

To do so, we assume that all stars \footnote{TWA~22 was not considered in this analysis, because previous studies \citep{Mamajek(2005),Barrado(2006),Teixeira(2009)} strongly suggest that this star is not a TWA member.}  in Table~3 with known trigonometric parallax are ``candidate members'' of the association. Then we come back to Eq.(\ref{eq.1})  and set $\sigma_{\mu_{\perp}}=0$  which is consistent with estimating an upper limit for the velocity dispersion of the association for those stars with known trigonometric parallax. The upper limit will allow us to recover as many members as possible in our upcoming CP analysis with a non-zero velocity dispersion value. A more refined velocity dispersion estimate will be discussed in Sect.~4.4 with our final sample of moving group members. The velocity dispersion $\sigma_{v}$ in Eq.~(1) that arises only from the perpendicular motion of stars is estimated iteratively, i.e., for each computed value of $\sigma_{v}$ we re-calculate the CP position. In the first iteration we use $\sigma_{v}=0$~km/s and the CP solution mentioned above. After a few iterations we converge towards $\sigma_{v}\simeq0.8\pm0.1$~km/s.

When we run the CP search method on the initial sample of TWA stars with $\sigma_{v}=0.8$~km/s we end up with a moving group of 30 stars that can be regarded as kinematic members of the association. The associated CP is located at  

\begin{center}
$(\alpha_{cp},\delta_{cp})=(100.1^{\circ},-27.1^{\circ})\pm(3.0^{\circ},1.3^{\circ})$,
\end{center}
with chi-squared statistics $\chi^{2}_{red}=1.2$ (i.e., $\chi^{2}/\nu=34.6/28$). We note that the addition of 12 stars (TWA~2, TWA~5, TWA~6, TWA~8A, TWA~8B, TWA~9A, TWA~9B, TWA~11A, TWA~11C, TWA~14, TWA~27, and TWA~28) to the 18 listed in the preceding paragraph shifts the CP position, but both results are still compatible within 1$\sigma$. Our solution is consistent with the CP derived by \citet{Mamajek(2005)}, $(\alpha_{cp},\delta_{cp})=(100.5^{\circ},-27.9^{\circ})\pm(5.0^{\circ},2.3^{\circ})$, using a different proper motion data set and CP method. However, ours is more precise which comes naturally from the proper motion data available at the moment.

%----------------------------------------------------------------------------------------------------------------
\subsection{Validation}

In the following we investigate via Monte Carlo simulations the validity of our CP solution derived with 30 moving group members (see Sect.~4.1). We construct 1000 synthetic samples of TWA by resampling the stellar proper motions from a Gaussian distribution where the mean and variance correspond to the proper motion and its uncertainty given in Table \ref{data}. For each simulated data set we run the CP search method (with $d=50$~pc and $\sigma_{v}=0.8$~km/s) and compute the CP location. The CP derived in Sect~4.1 is perfectly consistent with the centroid of simulated CPs (see Fig.~3) that is located at 
\begin{center}
$(\alpha_{cp},\delta_{cp})=(100.0^{\circ},-26.9^{\circ})\pm(2.3^{\circ},1.0^{\circ})$.
\end{center}
Thus, we conclude that our CP solution presented in Sect.~\ref{cp} is representative of the TWA moving group.

%FIGURE 3
\begin{figure}[!htp]
\begin{center}
\includegraphics[width=0.49\textwidth]{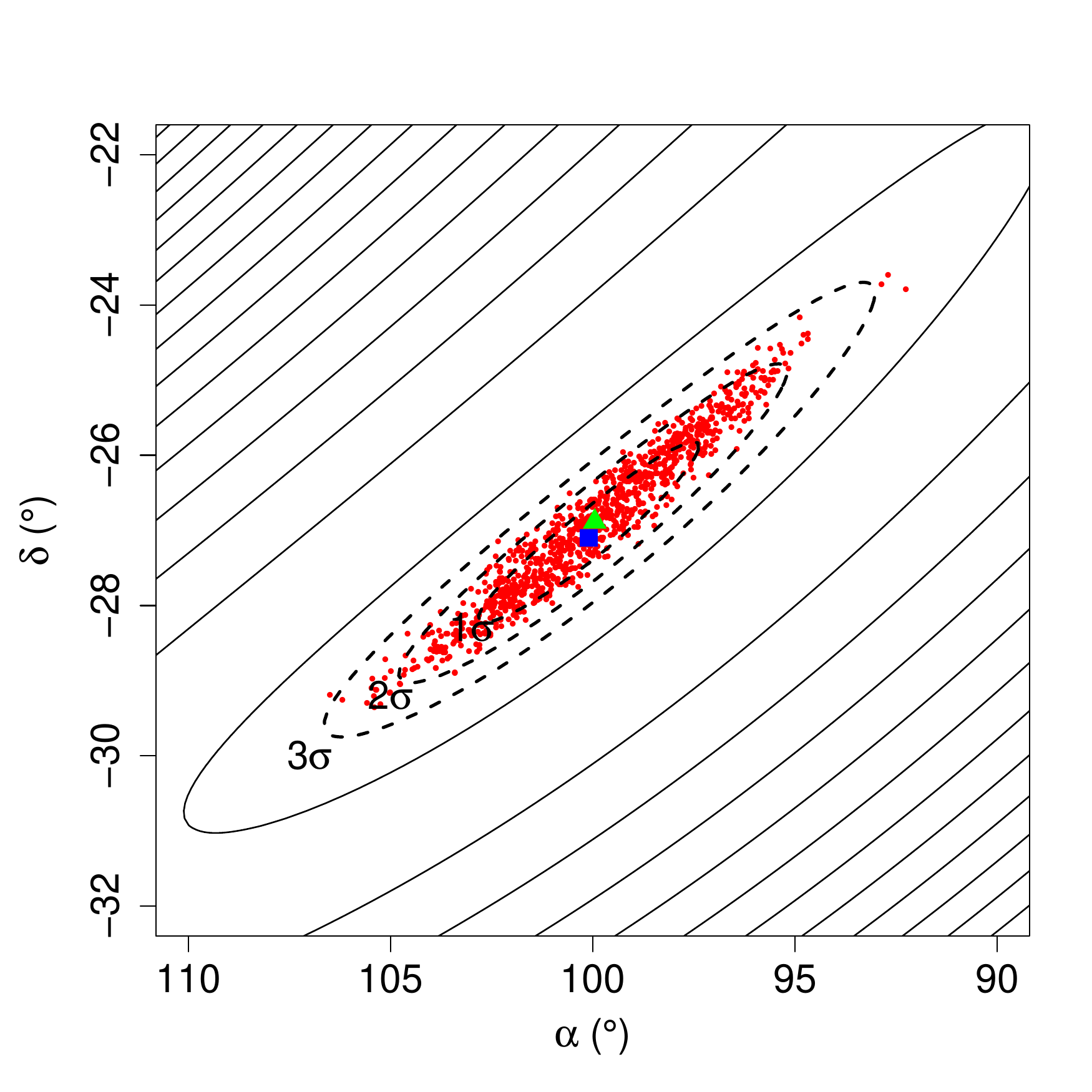}
\caption{\label{vector} Location of the simulated CPs for 1000 Monte Carlo realizations (red dots). The solid lines indicate the $X^{2}$ contours and the dashed lines denote the $1\sigma$, $2\sigma$ and $3\sigma$ confidence levels for the CP solution with 30 members presented in Sect.~4.1 (blue square). The green triangle indicates the centroid of simulated CPs.} 
\end{center}
\end{figure}

%----------------------------------------------------------------------------------------------------------------
\subsection{Kinematic parallaxes}

When a star belongs to a moving group its proper motion and radial velocity can be used
to determine its distance \citep[see e.g.][]{Mamajek(2005),Bertout(2006),Galli(2012)}. We use this approach to estimate the distance for 7 TWA members whose trigonometric parallax is not known in the literature. The kinematic parallax for each group member is given by
\begin{equation}
\pi_{kin}=\frac{A\, \mu_{\parallel}}{V_{r}\tan\lambda}\, ,
\end{equation}
where $\lambda$ is the angular distance from the CP to the star, $V_{r}$ is the radial velocity 
and $\mu_{\parallel}$ is the stellar proper motion component directed towards the CP (as defined in Sect.~4.1).

Among the 30 TWA members identified in our CP analysis only TWA~3, TWA~6, TWA~30A, TWA~30B, TWA~32, TWA~33 and TWA~34 do not have any trigonometric parallax measurement in the literature (and in this paper). We use the procedure described in the preceding paragraph and in Equation 4 to derive their kinematic parallax (equation 5 was used instead when radial velocity measurement was not available, see next paragraph); in Table~\ref{data} these values are marked with ``*'' to distinguish them from other  stars with trigonometric parallaxes.  This is the first determination of kinematic distance for TWA~30A, TWA~30B, TWA~32, TWA~33 and TWA~34, because they were not included in the previous CP analysis performed by \citet{Mamajek(2005)}.

For the specific case of TWA~3 we found five different radial velocity measurements \citep[see][]{Ramiro(1989),Reid(2003),Mamajek(2005),Torres(2006),Shkolnik(2011)}. TWA~3 is known to be a multiple system \citep[see e.g.][]{Schneider(2012a)}; this makes credible the various radial velocity values found in the literature. Since TWA~3 has not been resolved in current astrometric catalogs, the proper motion provided in Table~\ref{data} should be regarded as representative of the stellar system. On the other hand, TWA~33 and TWA~34 have been identified as TWA moving group members based on their proper motions and our CP analysis (see Sect.~4.1), but they do not have any measured radial velocity in the literature. In such cases, Eq.~(4) cannot be used to compute the stellar kinematic parallax. Thus, we use an alternative approach by assuming that those stars that belong to a moving group share the same space motion. In this case, the approximate parallax of the star is given by 
\begin{equation}
\pi_{app}=\frac{A\, \mu_{\parallel}}{V_{space}\sin\lambda}\, ,
\end{equation}
where the group spatial velocity is $V_{space}=21.5\pm 1.2$~km/s (to be discussed in Sect.~5). We use this approach to derive an approximate parallax for TWA~3, TWA~33 and TWA~34. We note that our distance estimates for TWA~33 ($d=53^{+3}_{-3}$~pc) and TWA~34 ($d=50_{-3}^{+3}$~pc) agree with the results presented by \citet{Schneider(2012b)}, but they are more precise which comes from the more precise proper motion data used in the present paper.

%---------------------------------------------------------------------------------------------------------------
\subsection{Membership analysis}

The final membership status of each star considered in this paper is given in Table~3 together with their astrometry. The 30 stars selected by the CP method in our analysis (see Sect.~4.1) are considered TWA moving group members and their membership status is indicated with `Y'. Our analysis confirms TWA~11A and TWA~11C as moving group members while the membership status of the TWA~11B component, whose proper motion is not known, cannot be investigated with the CP method. However, it seems unlikely that  TWA~11B is not co-moving with the other components of this multiple system, so we consider TWA~11B as a group member and mark its membership status with `Y'; we finally end up with a list of 31 kinematic members.

Among the rejected stars in our CP analysis we note that TWA~17, TWA~18, TWA~19A, TWA~19B, and TWA~24 have also been excluded as TWA members by \citet{Mamajek(2005)} because their distances inferred from the CP strategy are more consistent with the Lower Centaurus-Crux (LCC) subgroup of the Scorpius-Centaurus (Sco-Cen) association. That these stars are also rejected in our new and revised CP solution confirms the previous results. TWA~31 is also rejected by the CP search method, possibly due to its poor proper motion as compared to other TWA members. Its membership status should be re-discussed when more precise measurements become available. Our analysis also confirms that TWA~22 is not a group member as already suggested in previous studies \citep{Mamajek(2005),Barrado(2006),Teixeira(2009)}. All stars mentioned in this paragraph are therefore considered non-members of the association following our analysis, and their membership status in Table~3 is indicated with `N'.

On the other hand, TWA~10, TWA~16, TWA~20, and TWA~25 were considered TWA members in Mamajek's analysis, but are rejected by the CP search method in the present paper. By comparing the stellar proper motions used in both papers we conclude that those used here are more precise than the ones that were available when \citet{Mamajek(2005)} preformed his CP analysis. When we replace the proper motions given in Table~3 of the present paper for TWA~10, TWA~16, TWA~20, and TWA~25 by the ones listed in Table~1 of \citet{Mamajek(2005)}, and run the CP search method, we observe that these stars can be tolerated in our solution with negligible impact to the CP position. It seems that the large proper motion errors used in Mamajek's CP analysis have cast doubt on the derived $\mu_{\perp}$ allowing these stars to be considered TWA moving group members. Thus, given the more recent and precise proper motion measurements available now, such as the ones listed in Table~3, we cannot confirm TWA~10, TWA~16, TWA~20, and TWA~25 as TWA members based only on the CP search method. \citet{Galli(2012)} performed extensive simulations that convincingly demonstrate that more than 80$\%$ of all cluster members can be retrieved using the CP search method employed in this paper.  Although the fraction of cluster members identified with this technique is high, we may have missed a few group members. Additional information for TWA~10, TWA~16, TWA~20, and TWA~25, such as, parallax and radial velocity (see Table~3), spatial location (see Sect.~5), and isochronal age (see Sect.~6) suggest that they are consistent with membership in TWA. Thus, we retain these stars as possible TWA members, and mark their membership status in Table~3, column (13) with "Y?" to distinguish them from those group members that were directly identified by the CP search method.   

Once we have defined our final list of association members we are now in a better position to determine an accurate value for the velocity dispersion of the group. We estimate the intrinsic velocity dispersion of the association by the distribution of perpendicular velocities $V_{\perp}$ and their errors that arise from the $\mu_{\perp}$ proper motion components. Then, we search for the velocity dispersion $\sigma_{v}$ needed to force our result to $\chi^{2}_{red}=1$. We estimate the uncertainty of our result by defining the lower and upper limits for the velocity dispersion within an acceptable range of $\chi^{2}$ for a good fit \citep[see e.g.][]{Gould(2003)}. We find $\sigma_{v}=0.6^{+0.2}_{-0.1}$~km/s using all stars with known trigonometric parallax and marked with `Y' (and not 'Y?') in Table~3. Re-calculating the CP with $\sigma_{v}=0.6$~km/s has negligible effect in our solution. However, in Sect~4.3 we derived kinematic parallaxes for another 7 members of the association. Now that we have a distance estimate for all 30 moving group members identified in our CP analysis, there is no reason to limit the velocity dispersion analysis to only those 23~stars with known trigonometric parallax. Calculating the velocity dispersion with all 30 moving group members yields instead $\sigma_{v}=0.8^{+0.2}_{-0.1}$~km/s. We observe that our final velocity dispersion estimate confirms the results discussed in Sect.~4.1 and is in identical to the value of $\sigma_{v}=0.8^{+0.3}_{-0.2}$~km/s derived by \citet{Mamajek(2005)} in his analysis. \footnote{It is important to note that the upper limit of 0.8~km/s given in Sect.~4.1 refers \textit{only} to the sample of stars in Table~3 with known trigonometric parallax.}

%----------------------------------------------------------------------------------------------------------------
%					SPACE MOTION AND DYNAMICAL AGE 
%----------------------------------------------------------------------------------------------------------------
\section{Space motion and dynamical age\label{age}}

\citet{Makarov(2005)} and \citet{Ramiro(2006)} pioneered the idea of accurately deriving the dynamical age of the TWA.  \citet{Makarov(2005)} investigated the path of several probable members of the association and showed that 5 objects form a rapidly expanding association with an expansion age of 4.7 $\pm$ 0.6 My. \citet{Ramiro(2006)} showed that four \textit{Hipparcos} stars belonging to the association were converging back in time towards a minimum volume corresponding to the dynamical age of the association. This minimum volume was reached at an age of 8.3 Myr. Their different conclusions relied heavily on a small number of objects and on the accuracy of the data they used, in particular radial velocities. Since then, efforts have been dedicated to measurement of accurate parallaxes and proper motions of TWA stars \citep{Ducourant(2008),Teixeira(2008),Teixeira(2009),Weinberger(2013)} to secure the determination of the dynamical age of the association. 

In between, suspicion of a spread in the ages of the members was raised by several authors. \citet{Lawson(2005)} found evidence for two populations spatially and rotationally distinct, the TWA~1-13 group being younger ($\sim$10~Myr) and closer to the Sun while TWA~14-19 would constitute an older ($\sim$17~Myr) and more distant pre-main-sequence population rather associated with the LCC subgroup.   \citet{Barrado(2006)} revisited the age estimation of the association using various methods and notes that the TWA stars might not be coeval and that each age estimation method has its own limitations. 

Recently,\citet{Weinberger(2013)} published 14 parallaxes that greatly increased the number of available distances to TWA stars. They tried to derive a dynamical age for the association by using the astrometric data available but could not find any convergence back in time and concluded that TWA members have parallel motions. With a new set of available parallaxes and proper motions (from Table~\ref{data}), we are now in a situation to revisit the spatial motion of TWA stars and the dynamical age of the association.

%----------------------------------------------------------------------------------------------------------------
\subsection{Space Motion}

Among the 42~TWA stars (including resolved components) given in Table~\ref{data}, 31 stars have both known parallaxes (trigonometric or kinematic) and radial velocities. We computed the Galactic positions and velocities of these stars using the procedure described by \citet{Johnson(1987)}. These results are presented in Table~\ref{space}. The Galactic positions $(X,Y,Z)$ are given in a right-handed  coordinate system with origin at the Sun where $X$ points towards the Galactic center, $Y$ points in the direction of Galactic rotation and $Z$ points to the Galactic North pole. The stellar velocity projected on this XYZ grid defines the $(U,V,W)$ components of the Galactic space velocity. 

%A filamentary structure, appears clearly in the XY plane (as illustrated in Fig.~\ref{uvw}) with eventually two filaments merging in the top left part of the figure. 
Obviously, some objects exhibit motions or positions different from the rest. Are these differences a consequence of the history of these stars or of unreliable data? A  source of difficulty in such kinematic analysis is the quality and reliability of data, specially of radial velocities. Since many TWA stars are in double or multiple systems, the radial velocity of an individual star, depending on the separation of the components and the period of the system may change noticeably with time \citep{Makarov(2007)}.  This point is of great importance for the following section where we intend to trace stars back in time to find a common origin.

%TABLE 4
\begin{table*}[!htp]
\begin{center}
\caption{Galactic position and space velocity with respect to the Sun for the 31 stars from Table~\ref{data} with available parallaxes (trigonometric or kinematic) and radial velocities. 
\label{space}
}
\begin{tabular}{lcccrrr}

\hline
Star&$X$&$Y$&$Z$&$U$&$V$&$W$\\
&(pc)&(pc)&(pc)&(km/s)&(km/s)&(km/s)\\
\hline
    TWA1   &    7.5 &  -48.9 &   21.0 &  -12.1$\pm$   1.7 &  -18.5$\pm$   0.8 &   -6.4$\pm$   1.4\\
    TWA2   &    5.1 &  -36.5 &   19.5 &  -11.8$\pm$   2.6 &  -16.4$\pm$   1.5 &   -5.1$\pm$   2.0\\
    TWA3   &    6.5 &  -31.6 &   12.5 &  -11.9$\pm$   0.5 &  -15.1$\pm$   0.8 &   -5.6$\pm$   0.4\\
    TWA4   &    5.4 &  -36.9 &   24.9 &  -12.7$\pm$   1.5 &  -17.5$\pm$   1.0 &   -6.5$\pm$   1.2\\
    TWA5   &   11.3 &  -42.6 &   21.0 &  -10.8$\pm$   1.7 &  -20.2$\pm$   2.0 &   -4.3$\pm$   1.5\\
    TWA6   &   -1.4 &  -62.8 &   23.7 &  -10.9$\pm$   3.1 &  -21.0$\pm$   4.9 &   -8.4$\pm$   4.6\\
    TWA7   &    2.4 &  -31.9 &   12.9 &  -13.2$\pm$   1.0 &  -17.0$\pm$   0.5 &   -6.8$\pm$   0.8\\
    TWA8A  &    7.5 &  -35.1 &   23.1 &  -11.6$\pm$   1.1 &  -16.6$\pm$   0.9 &   -6.1$\pm$   1.0\\
    TWA8B  &    6.7 &  -31.8 &   20.9 &  -11.1$\pm$   1.0 &  -16.0$\pm$   0.7 &   -4.2$\pm$   0.7\\
    TWA9A  &   14.2 &  -40.4 &   18.8 &   -6.4$\pm$   1.2 &  -14.4$\pm$   0.8 &   -2.6$\pm$   0.9\\
    TWA9B  &   15.8 &  -45.0 &   21.0 &   -5.9$\pm$   0.8 &  -16.3$\pm$   1.8 &   -2.4$\pm$   0.9\\
    TWA10  &   28.5 &  -50.0 &   22.3 &  -10.9$\pm$   0.9 &  -17.7$\pm$   0.8 &   -7.0$\pm$   0.6\\
    TWA11A &   33.3 &  -58.4 &   28.4 &   -9.6$\pm$   1.2 &  -19.3$\pm$   1.9 &   -4.2$\pm$   1.0\\
    TWA12  &   14.9 &  -58.8 &   23.1 &  -13.3$\pm$   1.9 &  -20.0$\pm$   1.7 &   -5.5$\pm$   1.3\\
    TWA13A &   11.1 &  -49.3 &   23.1 &  -11.4$\pm$   0.8 &  -17.5$\pm$   0.7 &   -3.9$\pm$   0.6\\
    TWA13B &   11.9 &  -52.8 &   24.7 &  -12.8$\pm$   1.1 &  -18.9$\pm$   0.7 &   -3.9$\pm$   0.8\\
    TWA14  &   24.9 &  -89.9 &   23.5 &  -11.8$\pm$   2.0 &  -21.9$\pm$   2.1 &   -6.5$\pm$   1.4\\
    TWA15A &   53.2 &  -92.1 &   27.6 &  -10.3$\pm$   3.3 &  -20.4$\pm$   2.7 &   -3.7$\pm$   1.7\\
    TWA15B &   56.3 &  -97.5 &   29.2 &  -11.4$\pm$   3.4 &  -19.7$\pm$   2.7 &   -4.1$\pm$   2.0\\
    TWA16  &   37.2 &  -64.7 &   23.0 &   -9.2$\pm$   0.7 &  -18.2$\pm$   0.6 &   -5.6$\pm$   0.4\\
    TWA19A &   34.4 &  -83.0 &   18.6 &   -7.3$\pm$   2.0 &  -16.9$\pm$   3.5 &   -5.0$\pm$   1.3\\
    TWA20  &   36.4 &  -64.7 &   22.3 &  -14.1$\pm$   2.1 &  -20.8$\pm$   3.4 &   -9.2$\pm$   1.3\\
    TWA21  &    8.8 &  -49.6 &    2.8 &  -10.6$\pm$   1.0 &  -19.9$\pm$   0.8 &   -4.6$\pm$   0.5\\
    TWA22  &    3.5 &  -17.2 &    0.7 &   -8.2$\pm$   0.2 &  -15.9$\pm$   0.3 &   -9.0$\pm$   0.1\\
    TWA23  &   16.1 &  -39.2 &   23.7 &  -10.2$\pm$   1.2 &  -17.1$\pm$   1.3 &   -3.9$\pm$   0.9\\
    TWA25  &   21.5 &  -45.1 &   20.7 &  -10.5$\pm$   1.3 &  -18.6$\pm$   1.9 &   -5.6$\pm$   1.0\\
    TWA26  &    9.1 &  -32.3 &   18.2 &   -9.9$\pm$   0.7 &  -18.3$\pm$   1.7 &   -3.2$\pm$   1.0\\
    TWA27  &   19.5 &  -44.2 &   20.1 &   -8.0$\pm$   0.8 &  -18.2$\pm$   1.7 &   -3.6$\pm$   0.8\\
    TWA30A &   10.2 &  -43.2 &   25.1 &  -13.4$\pm$   2.2 &  -20.8$\pm$   1.9 &   -5.3$\pm$   1.8\\
    TWA30B &   10.3 &  -43.6 &   25.4 &  -12.1$\pm$   4.6 &  -20.6$\pm$   4.0 &   -6.2$\pm$   4.1\\
    TWA32  &   27.0 &  -52.8 &   33.3 &  -11.4$\pm$   1.6 &  -18.0$\pm$   1.4 &   -4.7$\pm$   1.1\\
\hline
\end{tabular}
\end{center}
\end{table*}

%----------------------------------------------------------------------------------------------------------------
\subsection{Traceback age of TWA}\label{trace}

To determine the traceback age of the association, we considered the 25 stars, kinematic members as determined in Sect. \ref{cp} (status='Y' in Table \ref{data}) that have radial velocity and parallax measurements (TWA~1, TWA~2, TWA~3, TWA~4, TWA~5, TWA~6, TWA~7, TWA~8A, TWA~8B, TWA~9A, TWA~9B, TWA~11A, TWA~12, TWA~13A, TWA~13B, TWA~14, TWA~15A, TWA~15B, TWA~21, TWA~23, TWA~26, TWA~27, TWA~30A, TWA~30B and TWA~32). 

We then considered the present day positions of these objects and computed their location backwards in time with a step of 0.1~Myr for a period of 20~Myr. To characterize the extent of the association we computed at each epoch the mean coordinates ($\bar{X},\bar{Y},\bar{Z}$) of the group and their associated standard deviation about the mean ($\sigma_{X},\sigma_{Y},\sigma_{Z}$). We defined the typical radius of the association as : $rad =  \frac{1}{3}( \sigma_{X} + \sigma_{Y} + \sigma_{Z})$ and searched for the epoch minimizing this quantity. This radius should be representative of the global spread of stars around the mean.

The examination of the evolution with time of the distance of each object from the mean of the group revealed that several objects systematically drifted away from the center of the association back in time. This is the case for TWA~5, TWA~6, TWA~9A, TWA~9B, TWA~21, TWA~26, TWA~27, TWA~30A and TWA~30B.  Several reasons may explain this behavior, such as corrupted data (most probably), contamination by non members or non coevality with others stars. 

Eliminating these 9 discrepant stars leaves us with a sample of 16 stars (TWA~1, TWA~2, TWA~3, TWA~4, TWA~7, TWA~8A, TWA~8B, TWA~11A, TWA~12, TWA~13A, TWA~13B, TWA~14, TWA~15A, TWA~15B, TWA~23 and TWA~32). We will designate these converging members, hereafter as \textit{traceback core} stars. Their participation to the determination of the trace-back age is indicated in Table~3, column (14) : "+" for traceback core stars, "r" for objects rejected because they were systematically drifting away from TWA center and "--" for non members as defined by CP analysis or when data were missing for space velocity calculation.

\begin{figure}[!htp]
\begin{center}
\includegraphics[width=0.49\textwidth]{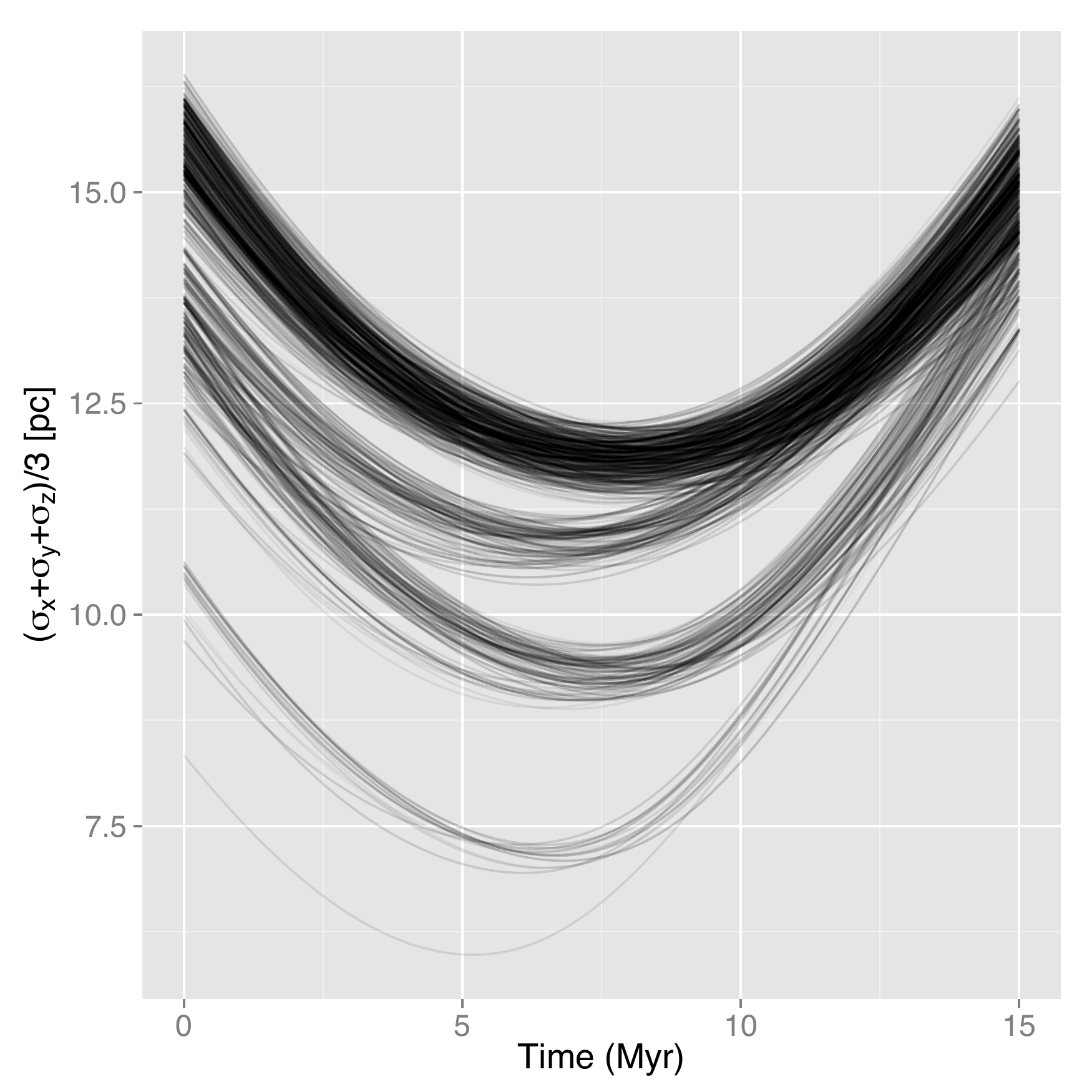}
\caption{Evolution back in time of the radius $rad$ of the 2000 resampled lists of 13 TWA core-stars. 
%Minimum radius are indicated
\label{radius}}
\end{center}
\end{figure}

\begin{figure}[!htp]
\begin{center}
\includegraphics[width=0.49\textwidth]{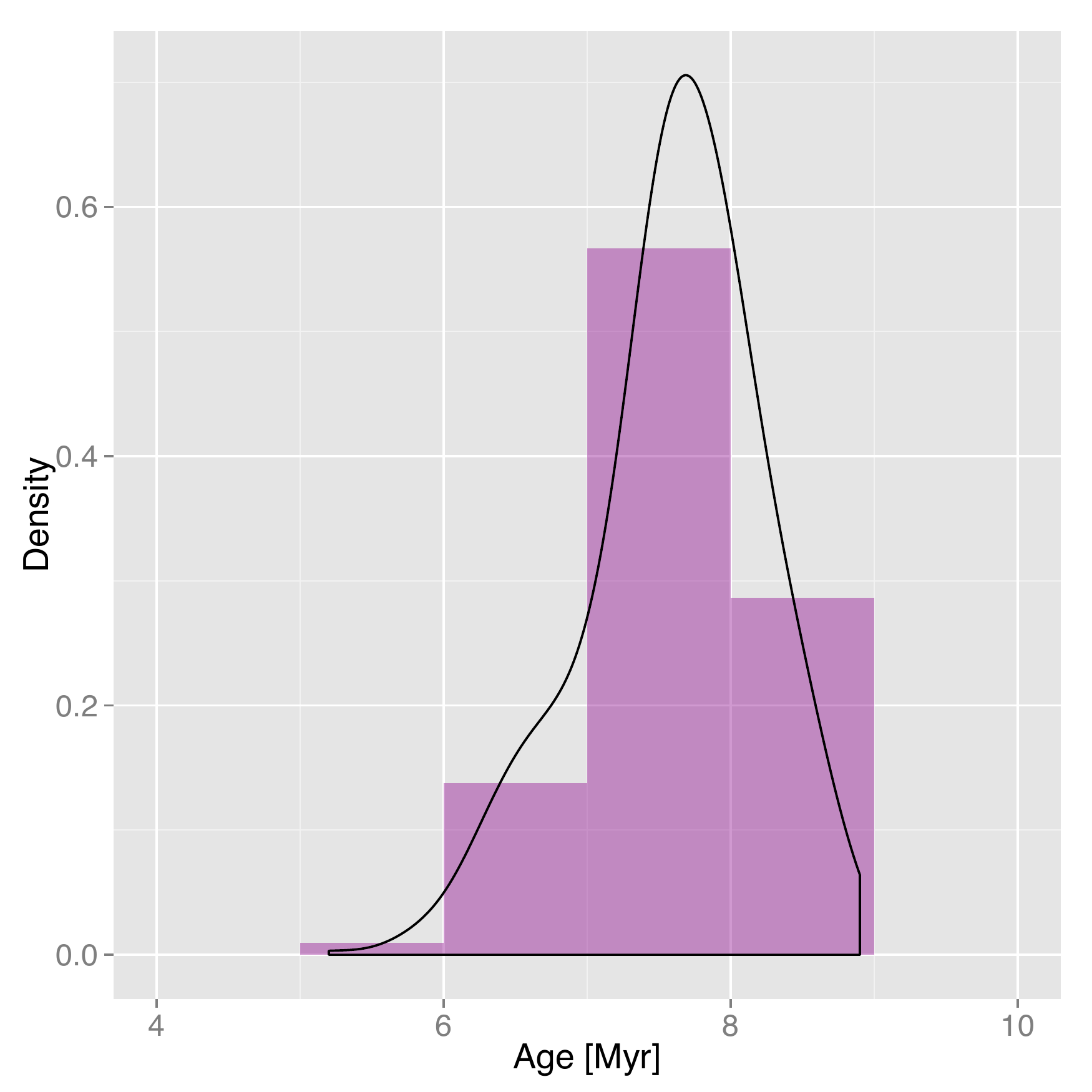}
\caption{Distribution of epochs corresponding to the minimum radius of the 2000 resampled lists of 13 TWA core-stars as indicated by a histogram (in purple) and a kernel density estimator (in black). The mean of the distribution and dispersion about the mean indicate a trace-back age for the association of $t\simeq-7.5 \pm0.7$~Myr.
%The mean of the distribution and the dispersion about the mean designate the trace-back age of TWA :  $t\simeq-7.5 \pm0.7$~Myr.
% Histogram of epochs
%$t\simeq-7.5 \pm0.7$~Myr.
\label{time}}
\end{center}
\end{figure}

Assuming that our sample of stars may be contaminated by non-members and to get rid of the particular influence of each star we applied a Jackknife resampling technique to our list of 16 core-stars eliminating randomly 3 objects (20 $\%$ of the sample). We generated 2000 random lists of 13 stars taken from our 16 core-stars and calculated for each the epoch of convergence when $rad$ is minimum. Each possible configuration of 13 stars drawn from 16 was represented about 3 or 4 times in our trials. The mean of the epochs obtained and the dispersion about the mean correspond to the back-track age of TWA $t\simeq-7.5 \pm0.7$~Myr. We present in Fig.~\ref{radius} the evolution of the radius of the association $rad$  as function of time for the 2000 configurations and in Fig.~\ref{time} the repartition of the epochs corresponding to the minimum radius.

%FIGURE 4

%FIGURE 5

One notices several groups of curves in this figure. The upper one contains the most configurations. The lowest ones correspond to samples where TWA~15A or/and TWA~15B are absent.  These stars have the smallest parallaxes of any in Table \ref{data}, so that their non-inclusion naturally diminishes the radius of the association.  One could question whether these stars are members of TWA and instead place them in Lower Centaurus-Crux (LCC) that is located behind the TWA. However, TWA 15A and 15B are consistent with TWA membership according to our CP analysis and their evolutionary ages derived in Section 6 are more consistent with the mean evolutionary age of TWA than with that of LCC (17 Myr, \citet{Chen(2011)}; even if LCC is not as old as 17 Myr, it is older than TWA (Figure 2 in \citet{Song(2012)}). Moreover a chi square test applied to the UVW velocities of these two objects to test their ad-equation to the mean velocities of TWA or LCC systematically better agrees with a kinematic membership to TWA. Therefore, we retain TWA15A and B in our list of core trace-back stars. A trace-back strategy applied to the list excluding TWA~15A and TWA~15B leads to a traceback age of -6.3+/-0,8 Myr.

The weighted mean UVW velocities of the 16 \textit{traceback core} stars are $(-11.7,-17.3,-5.0)\pm(0.9, 1.3, 1.0)$~km/s. These values are in good agreement with various published values \citep[e.g.][]{Torres(2008),Weinberger(2013)} and confirms that our sample of \textit{traceback core} stars consists of genuine TWA members. These converging members lie at a mean distance d~$\simeq$61 pc (ranging from 34.6 to 116.3 pc).

The age derived in this work can be compared to similar investigations via backtracking. \cite{Makarov(2005)} derived an expansion age of 4.7 Myr using Hipparcos data for TWA 1, 4, 11 and two other young nearby stars (HD 139084 and HD 220476). \cite{Ramiro(2006)} derived an expansion age of 8.3$\pm$0.8 Myr using Hipparcos data for TWA 1, 4, and 11 and 19. These two determinations based essentially on the same astrometric data (Hipparcos) differ by the selection of stars (inclusion or not of TWA 19) and inclusion of two stars which are probably not members. Using new parallaxes and available UCAC3 proper motions \cite{Weinberger(2013)} could not observe any convergence of the TWA group back in time. In the present work we used an updated database including the UCAC4 proper motions and new trigonometric parallaxes and observed a convergence for a restricted list of 16 stars. Indeed the data and the number of objects used to derive the age have a major impact on the result. Finally the reality of the membership of the objects used for the backtracking is essential. Moreover if some TWA groups of stars are not coeval then the task of finding an expansion age become even more complex. In this work we selected a subsample of probable coeval members of TWA and derived a traceback age for the association. This result will be considered in the next section via evaluation of individual star ages by the study of an HR diagram.

%%----------------------------------------------------------------------------------------------------------------
%%							HR-DIAGRAM 
%%----------------------------------------------------------------------------------------------------------------
\section{HR-Diagram}\label{HR}

In the following, we construct an HR-diagram of the TWA and determine the main physical parameters of individual stars using available photometric and spectroscopic information. Our analysis is obviously restricted to those stars marked with `Y' and `Y?' in Table~3 with known parallaxes and photometry. For a few systems that were resolved by high angular resolution photometric observations but with unresolved astrometry we assume the same parallax value for the various components. 

\citet{Weinberger(2013)} derived the age of 29 TWA stars using their new parallax results and $H$ band photometry from the 2MASS catalog. No attempt was made in their work to correct the computed absolute magnitudes for extinction. In this paper we take into account the effect of stellar extinction due to circumstellar material that may be present in some association members \citep[see][]{Schneider(2012a)}. We compute the visual extinction $A_{V}$ from the color excess in $(B-V)$, $(V-I_{C})$, $(V-J)$, $(V-H)$ and $(V-K)$ following the procedure described by \citet{Pecaut(2012)}. We also used $(J-K)$ from 2MASS photometry when optical colors were not available \citep[see e.g. Sect.~4.2 of ][]{Kenyon(1995)}. Most stars in our sample exhibit low extinction, so we set the extinction to zero when derived  estimates of $A_{V}$ yield a non-physical negative value that one might attribute to photometric errors. We adopt a total to selective extinction ratio of $R_{V}=3.1$ and took the weighted mean of all significant non-zero $A_{V}$ values as our final result. The formal uncertainties in our $A_{V}$ estimates that come from photometric errors are better than $0.1$~mag. Our results are given in Table~5. We use the value of $A_{H}/A_{V}=0.15$ from \citet{Cieza(2005)} to compute $A_{H}$ for each star and correct the stellar luminosities (see below) from stellar extinction. The $A_{H}$ values typically ranged from 0 to 0.05~mag.  

The stellar luminosities were derived from the $H$ flux to better compare our results with \citet{Weinberger(2013)}. However, to ensure the quality of our results we built a control sample by re-calculating the stellar luminosities (and the stellar parameters) from the 2MASS $J$ flux. Since both procedures yield equivalent results within 8\% and for clarity of presentation we have chosen to present in Table~\ref{mass_age} only the stellar luminosities that result from the $H$ flux. The luminosity error budget takes into account photometric errors and the parallax uncertainty. We used the intrinsic colors, temperatures and bolometric corrections for 5-30~Myr pre-main sequence stars as given in Table~6 of \citet{Pecaut(2013)}.\footnote{Because TWA~11A is an A0 star \citep[see e.g.][]{Schneider(2012a)}, we used the intrinsic colors for dwarf stars given in Table~4 of \citet{Pecaut(2013)} to compute its luminosity. }

%TABLE 5
\begin{table*}[!htp]
\begin{center}
\caption{\label{mass_age}
Physical parameters derived for TWA stars with known distance.}
\resizebox{19cm}{!} {
\begin{tabular}{lcccccccccccccc}
\hline
&&&&&&\multicolumn{2}{c}{\citet{Siess(2000)}}&\multicolumn{2}{c}{\citet{Baraffe(1998)}}&\\
\hline
Star&ST&$T_{eff}$&$A_{V}$&$H_{abs}$&$\log(L/L_{\odot})$&$\log(M/M_{\odot})$&$\log t$&$\log(M/M_{\odot})$&$\log t$&Ref.\\
&&(K)&(mag)&(mag)&&&(t in Myr)&&(t in Myr)&\\
\hline

TWA 1	& 	K8	& 	3940	& 	0.00	&$ 	3.91	\pm 	0.25	$&$ 	-0.58	\pm 	0.12	$& 	$	-0.16	\pm 	0.01	$&$ 	6.90	\pm 	0.19	$&$ 	-0.10	\pm 	0.03	$&$ 	7.09	\pm 	0.21	$& 	1,2,3,4	\\
TWA 2A	& 	M0.5	& 	3704	& 	0.33	&$ 	4.22	\pm 	0.44	$&$ 	-0.75	\pm 	0.20	$& 	$	-0.34	\pm 	0.01	$&$ 	6.81	\pm 	0.26	$&$ 	-0.23	\pm 	0.08	$&$ 	7.02	\pm 	0.29	$& 	1,3,6,10	\\
TWA 2B	& 	M2	& 	3490	& 	0.00	&$ 	5.01	\pm 	0.44	$&$ 	-1.09	\pm 	0.20	$& 	$	-0.50	\pm 	0.03	$&$ 	6.91	\pm 	0.24	$&$ 	-0.39	\pm 	0.06	$&$ 	7.13	\pm 	0.27	$& 	1,3,5,6	\\
TWA 3A	& 	M4	& 	3160	& 	0.00	&$ 	4.79	\pm 	0.13	$&$ 	-1.03	\pm 	0.07	$& 	$	-0.75	\pm 	0.02	$&$ 	6.60	\pm 	0.05	$&$ 	-0.74	\pm 	0.04	$&$ 	6.38	\pm 	0.07	$& 	3,5,6	\\
TWA 3B	& 	M4	& 	3160	& 	0.00	&$ 	5.41	\pm 	0.14	$&$ 	-1.28	\pm 	0.08	$& 	$	-0.81	\pm 	0.02	$&$ 	6.78	\pm 	0.06	$&$ 	-0.81	\pm 	0.03	$&$ 	6.61	\pm 	0.06	$& 	3,5,6	\\
TWA 4A	& 	K5	& 	4140	& 	0.00	&$ 	3.31	\pm 	0.24	$&$ 	-0.31	\pm 	0.12	$& 	$	-0.06	\pm 	0.01	$&$ 	6.73	\pm 	0.20	$&$ 	-0.04	\pm 	0.01	$&$ 	6.81	\pm 	0.22	$& 	5,7,8	\\
TWA 4B	& 	K7,M1	& 	3970	& 	0.00	&$ 	3.32	\pm 	0.24	$&$ 	-0.34	\pm 	0.12	$& 	$	-0.16	\pm 	0.01	$&$ 	6.56	\pm 	0.18	$&$ 	-0.10	\pm 	0.03	$&$ 	6.70	\pm 	0.20	$& 	5,7,8	\\
TWA 5Aa	& 	M2	& 	3490	& 	0.00	&$ 	4.25	\pm 	0.26	$&$ 	-0.79	\pm 	0.12	$& 	$	-0.46	\pm 	0.01	$&$ 	6.55	\pm 	0.10	$&$ 	-0.40	\pm 	0.06	$&$ 	6.70	\pm 	0.17	$& 	1,3,5,9	\\
TWA 5Ab	& 	M2	& 	3490	& 	0.00	&$ 	4.35	\pm 	0.26	$&$ 	-0.83	\pm 	0.12	$& 	$	-0.47	\pm 	0.01	$&$ 	6.60	\pm 	0.14	$&$ 	-0.39	\pm 	0.05	$&$ 	6.75	\pm 	0.18	$& 	1,3,5,9	\\
TWA 6	& 	M0	& 	3770	& 	0.00	&$ 	4.05	\pm 	0.64	$&$ 	-0.67	\pm 	0.28	$& 	$	-0.29	\pm 	0.01	$&$ 	6.72	\pm 	0.39	$&$ 	-0.19	\pm 	0.06	$&$ 	7.00	\pm 	0.43	$& 	1,3,4	\\
TWA 7	& 	M3	& 	3360	& 	0.00	&$ 	4.44	\pm 	0.16	$&$ 	-0.87	\pm 	0.07	$& 	$	-0.56	\pm 	0.01	$&$ 	6.61	\pm 	0.08	$&$ 	-0.54	\pm 	0.05	$&$ 	6.55	\pm 	0.09	$& 	1,3,4	\\
TWA 8A	& 	M3	& 	3360	& 	0.00	&$ 	4.57	\pm 	0.21	$&$ 	-0.93	\pm 	0.11	$& 	$	-0.56	\pm 	0.02	$&$ 	6.63	\pm 	0.12	$&$ 	-0.55	\pm 	0.05	$&$ 	6.62	\pm 	0.15	$& 	1,3,10	\\
TWA 9A	& 	K7	& 	3970	& 	0.00	&$ 	4.60	\pm 	0.27	$&$ 	-0.86	\pm 	0.14	$& 	$	-0.16	\pm 	0.03	$&$ 	7.38	\pm 	0.22	$&$ 	-0.15	\pm 	0.08	$&$ 	7.58	\pm 	0.21	$& 	1,2,3,10	\\
TWA 9B	& 	M3.5	& 	3255	& 	0.00	&$ 	5.83	\pm 	0.16	$&$ 	-1.44	\pm 	0.09	$& 	$	-0.74	\pm 	0.01	$&$ 	7.02	\pm 	0.09	$&$ 	-0.72	\pm 	0.03	$&$ 	6.97	\pm 	0.11	$& 	1,3,10	\\
TWA 10	& 	M2	& 	3490	& 	0.36	&$ 	4.47	\pm 	0.14	$&$ 	-0.88	\pm 	0.07	$& 	$	-0.47	\pm 	0.01	$&$ 	6.66	\pm 	0.08	$&$ 	-0.39	\pm 	0.05	$&$ 	6.82	\pm 	0.11	$& 	1,3,4	\\
TWA 11A	& 	A0	& 	9700	& 	0.00	&$ 	1.48	\pm 	0.06	$&$ 	1.40	\pm 	0.03	$& 	$	+0.35	\pm 	0.01	$&$ 	6.80	\pm 	0.06$&................&................& 2,4,5	\\
TWA 11B	& 	M2	& 	3490	& 	0.00	&$ 	4.30	\pm 	0.14	$&$ 	-0.81	\pm 	0.08	$& 	$	-0.46	\pm 	0.01	$&$ 	6.58	\pm 	0.09	$&$ 	-0.40	\pm 	0.05	$&$ 	6.73	\pm 	0.10	$& 	5,10	\\
TWA 11C	& 	M4.5	& 	3001	& 	0.00	&$ 	5.03	\pm 	0.08	$&$ 	-1.13	\pm 	0.04	$& 	$	-0.95	\pm 	0.01	$&$ 	6.60	\pm 	0.02	$&$ 	-1.00	\pm 	0.02	$&$ 	6.18	\pm 	0.08	$& 	1,4	\\
TWA 12	& 	M2	& 	3490	& 	0.27	&$ 	4.23	\pm 	0.24	$&$ 	-0.78	\pm 	0.11	$& 	$	-0.46	\pm 	0.01	$&$ 	6.55	\pm 	0.12	$&$ 	-0.40	\pm 	0.06	$&$ 	6.69	\pm 	0.15	$& 	1,3,4	\\
TWA 13A	& 	M1	& 	3630	& 	0.00	&$ 	4.00	\pm 	0.11	$&$ 	-0.67	\pm 	0.06	$& 	$	-0.38	\pm 	0.01	$&$ 	6.54	\pm 	0.08	$&$ 	-0.27	\pm 	0.05	$&$ 	6.80	\pm 	0.08	$& 	1,3,4	\\
TWA 13B	& 	M1	& 	3630	& 	0.31	&$ 	3.76	\pm 	0.11	$&$ 	-0.57	\pm 	0.06	$& 	$	-0.38	\pm 	0.01	$&$ 	6.42	\pm 	0.07	$&$ 	-0.28	\pm 	0.05	$&$ 	6.67	\pm 	0.09	$& 	1,3,4	\\
TWA 14	& 	M0	& 	3770	& 	0.00	&$ 	4.57	\pm 	0.26	$&$ 	-0.88	\pm 	0.12	$& 	$	-0.30	\pm 	0.01	$&$ 	7.02	\pm 	0.19	$&$ 	-0.21	\pm 	0.05	$&$ 	7.33	\pm 	0.19	$& 	4,5,11	\\
TWA 15A	& 	M1.5	& 	3562	& 	0.00	&$ 	4.73	\pm 	0.41	$&$ 	-0.97	\pm 	0.18	$& 	$	-0.44	\pm 	0.01	$&$ 	6.84	\pm 	0.22	$&$ 	-0.32	\pm 	0.06	$&$ 	7.10	\pm 	0.25	$& 	4,5,12	\\
TWA 15B	& 	M2	& 	3490	& 	0.00	&$ 	4.49	\pm 	0.41	$&$ 	-0.88	\pm 	0.18	$& 	$	-0.48	\pm 	0.02	$&$ 	6.73	\pm 	0.20	$&$ 	-0.39	\pm 	0.05	$&$ 	6.83	\pm 	0.26	$& 	4,5,12	\\
TWA 16	& 	M2	& 	3490	& 	0.00	&$ 	4.56	\pm 	0.10	$&$ 	-0.91	\pm 	0.06	$& 	$	-0.47	\pm 	0.01	$&$ 	6.70	\pm 	0.06	$&$ 	-0.39	\pm 	0.05	$&$ 	6.88	\pm 	0.09	$& 	1,4,13	\\
TWA 20	& 	M3	& 	3360	& 	0.00	&$ 	5.01	\pm 	0.13	$&$ 	-1.10	\pm 	0.07	$& 	$	-0.59	\pm 	0.01	$&$ 	6.81	\pm 	0.07	$&$ 	-0.56	\pm 	0.05	$&$ 	6.86	\pm 	0.10	$& 	1,4	\\
TWA 21	& 	K3	& 	4550	& 	0.00	&$ 	3.84	\pm 	0.16	$&$ 	-0.43	\pm 	0.08	$& 	$	-0.04	\pm 	0.03	$&$ 	7.41	\pm 	0.10	$&$ 	-0.05	\pm 	0.05	$&$ 	7.40	\pm 	0.12	$& 	1,4	\\
TWA 23	& 	M1	& 	3630	& 	0.00	&$ 	5.35	\pm 	0.19	$&$ 	-1.21	\pm 	0.09	$& 	$	-0.41	\pm 	0.01	$&$ 	7.25	\pm 	0.13	$&$ 	-0.28	\pm 	0.06	$&$ 	7.59	\pm 	0.14	$& 	1,4	\\
TWA 25	& 	K9	& 	3880	& 	0.33	&$ 	3.79	\pm 	0.15	$&$ 	-0.55	\pm 	0.08	$& 	$	-0.21	\pm 	0.01	$&$ 	6.72	\pm 	0.11	$&$ 	-0.12	\pm 	0.03	$&$ 	6.98	\pm 	0.12	$& 	1,3,4	\\

\hline
\end{tabular}
}
\end{center}

\tablefoot{
We provide for each star the spectral type, temperature, visual extinction, absolute magnitude from H flux, luminosity, mass and age derived from \citet{Siess(2000)} and \citet{Baraffe(1998)} models, and the sources of photometric/spectroscopic information. Note that the photometry of TWA~2, TWA~3, TWA~4, TWA~5, TWA~8, TWA~9, and TWA~16 are corrected for binarity with the information provided in the corresponding references (last column). TWA~14, TWA~20, and TWA~23 are corrected for binarity assuming that the components are equal brightness \citep[as done by][]{Weinberger(2013)}. Note that our age and mass estimates for TWA~11C derived from the \citet{Baraffe(1998)} tracks and isochrones are only indicative due to the grid limits of this model and the star position in the HR-diagram (see Fig.~6). 
}

\tablebib{
(1)~Pecaut \& Mamajek (2013);
(2)~ \textit{Hipparcos} \citep{HIP97};
(3)~\citet{Torres(2006)};
(4)~2MASS \citep{2MASS};
(5)~\citet{Schneider(2012a)};
(6)~\citet{Brandeker(2003)};
(7)~\citet{Fernandez(2008)};
(8)~\citet{Prato(2001)};
(9)~\citet{Konopacky(2007)};
(10)~\citet{Webb(1999)};
(11)~\citet{Messina(2010)};
(12)~\citet{Barrado(2006)};
(13)~\citet{Zuckerman(2001)};
(14)~\citet{Teixeira(2008)};
(15)~\citet{Looper(2010a)}.
\vspace{1cm}
}
\end{table*}
%\bigskip

\begin{figure*}[!htp]
\begin{center}
\includegraphics[width=0.6\textwidth]{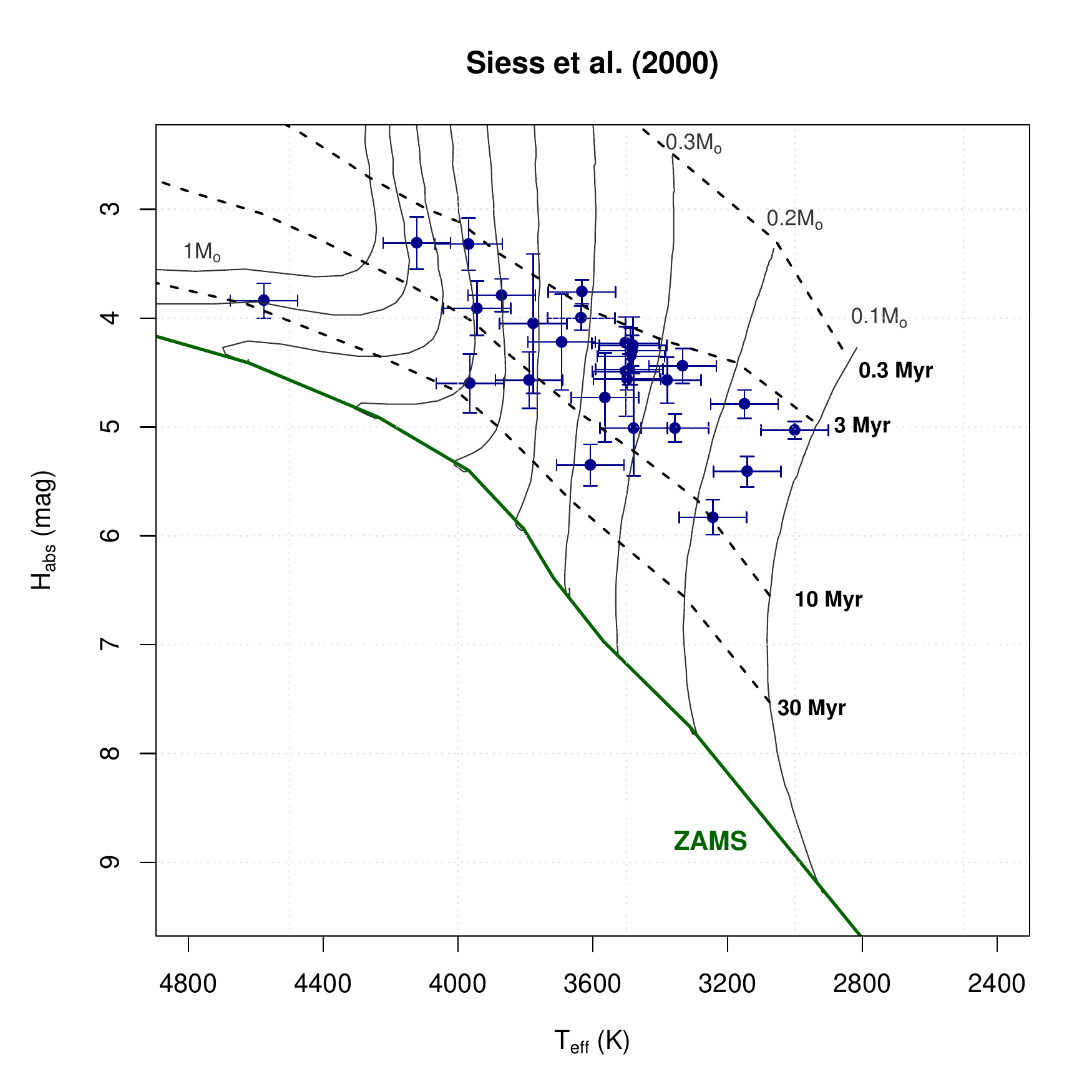}
\includegraphics[width=0.6\textwidth]{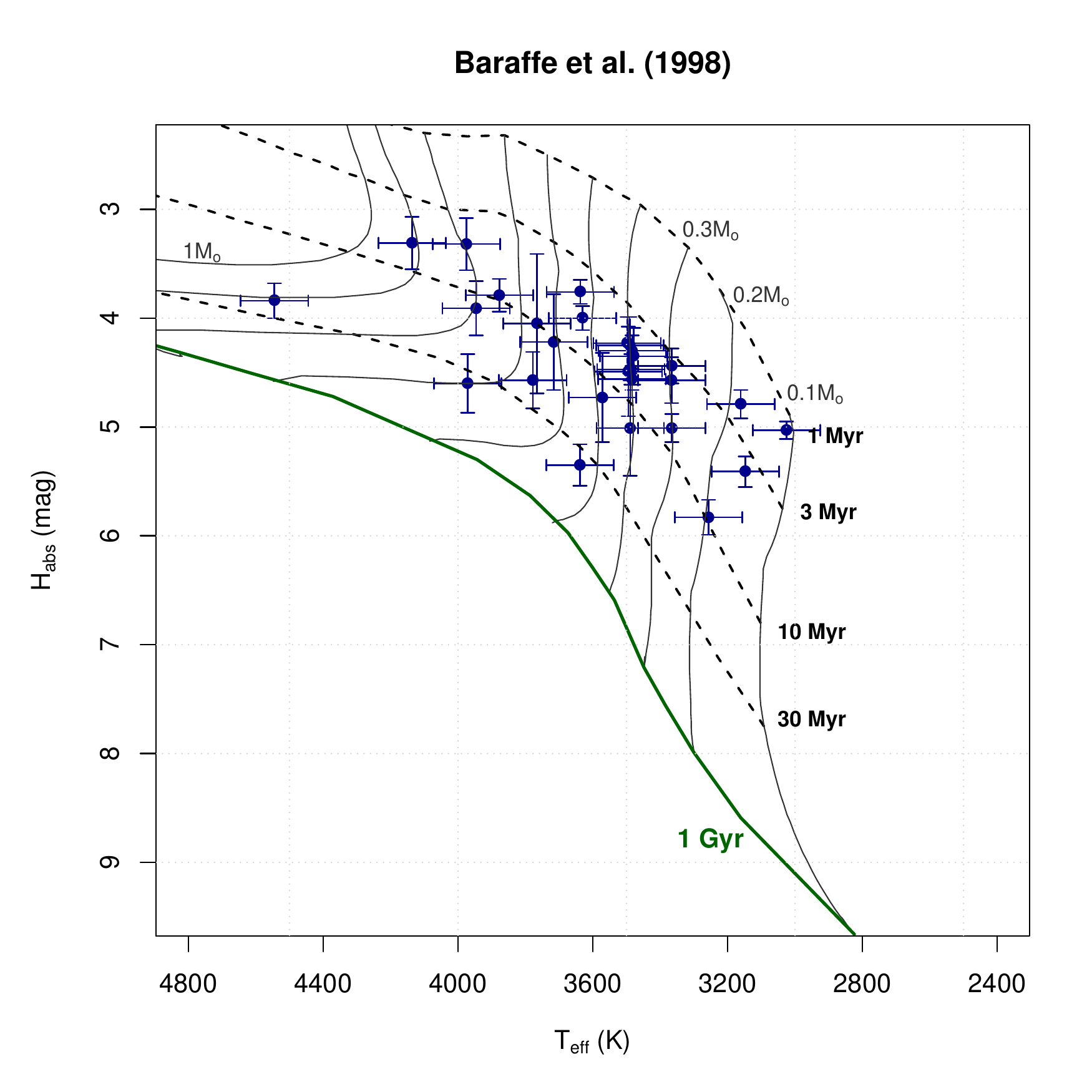}
\caption{HR-diagram of the TWA stars listed in Table~5 with the grid of evolutionary models computed by \citet{Siess(2000)} with metallicity $Z=0.02$ \textit{(upper panel)} and \citet{Baraffe(1998)} with $Y=0.282$ and mixing length parameter of 1.9 \textit{(lower panel)}. The solid and dashed lines are, respectively, evolutionary tracks from $0.1M_{\odot}$ to $1M_{\odot}$ with mass increment of $0.1M_{\odot}$ and theoretical isochrones with the ages indicated in the figure. The green line indicates the zero-age main sequence (ZAMS) with the \citet{Siess(2000)} models, and the 1~Gyr isochrone computed by \citet{Baraffe(1998)}. We assume a $\pm100$~K uncertainty for all spectral types. 
\label{HRD}}
\end{center}
\end{figure*}

\begin{figure*}[!]
\begin{center}
\includegraphics[width=0.45\textwidth]{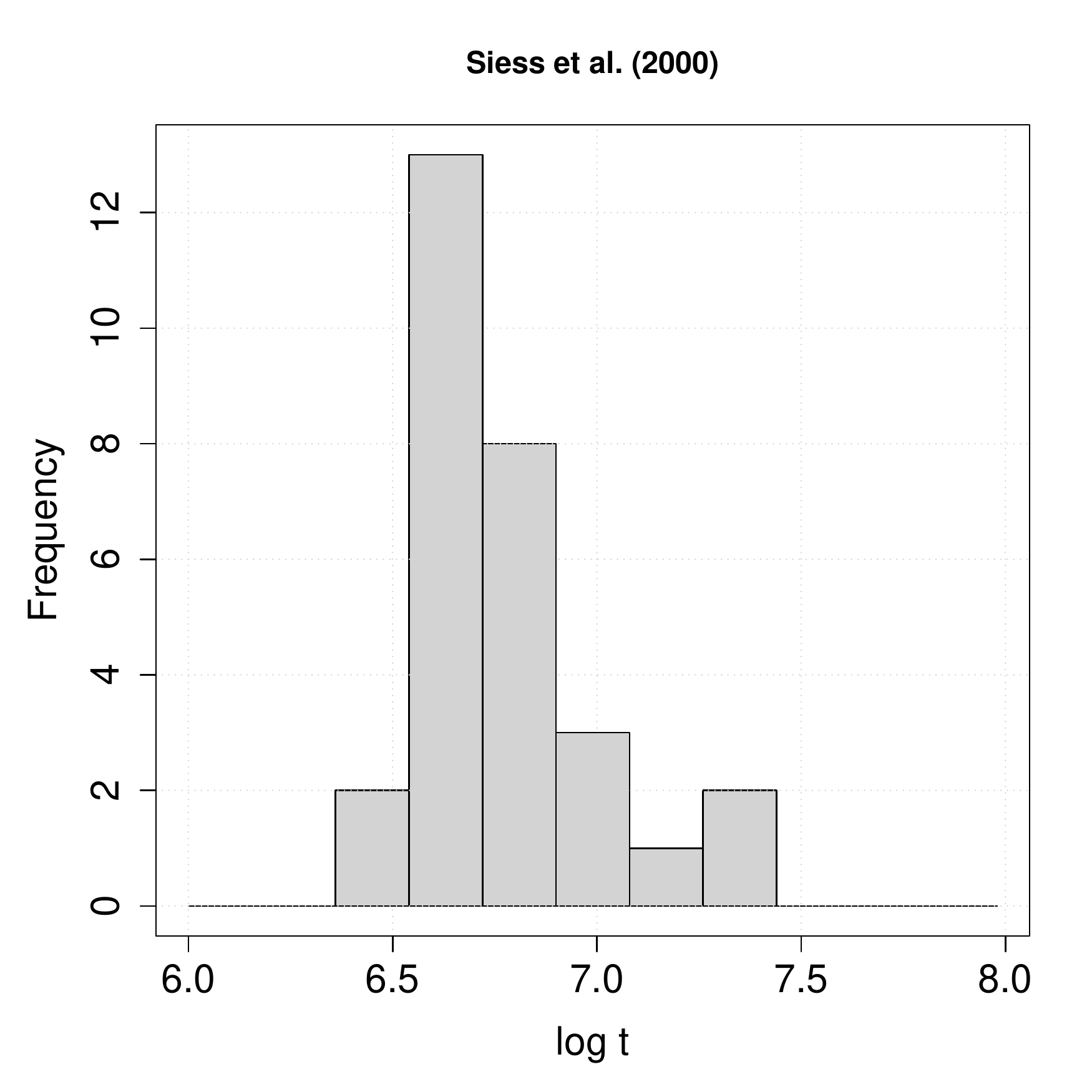}
\includegraphics[width=0.45\textwidth]{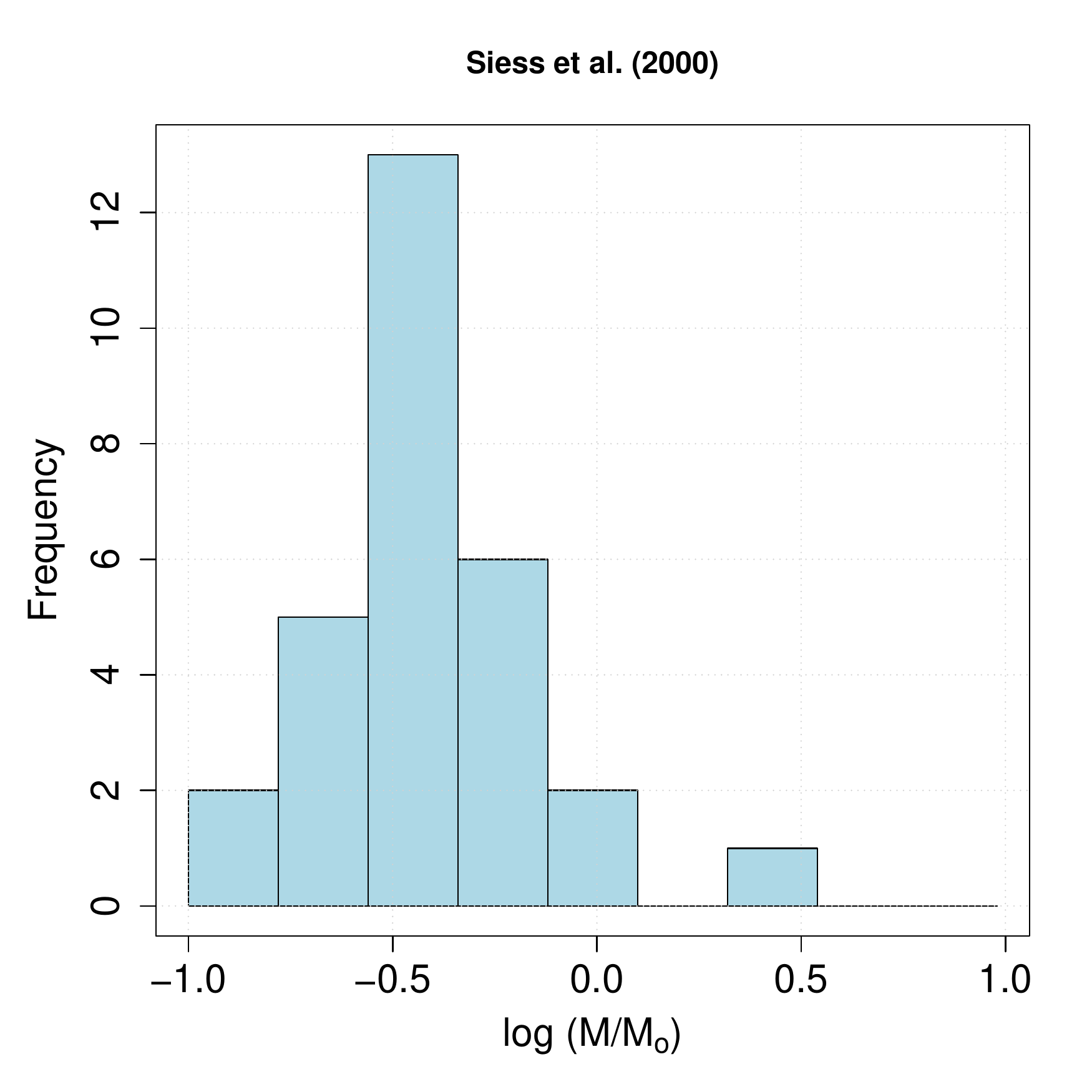}
\includegraphics[width=0.45\textwidth]{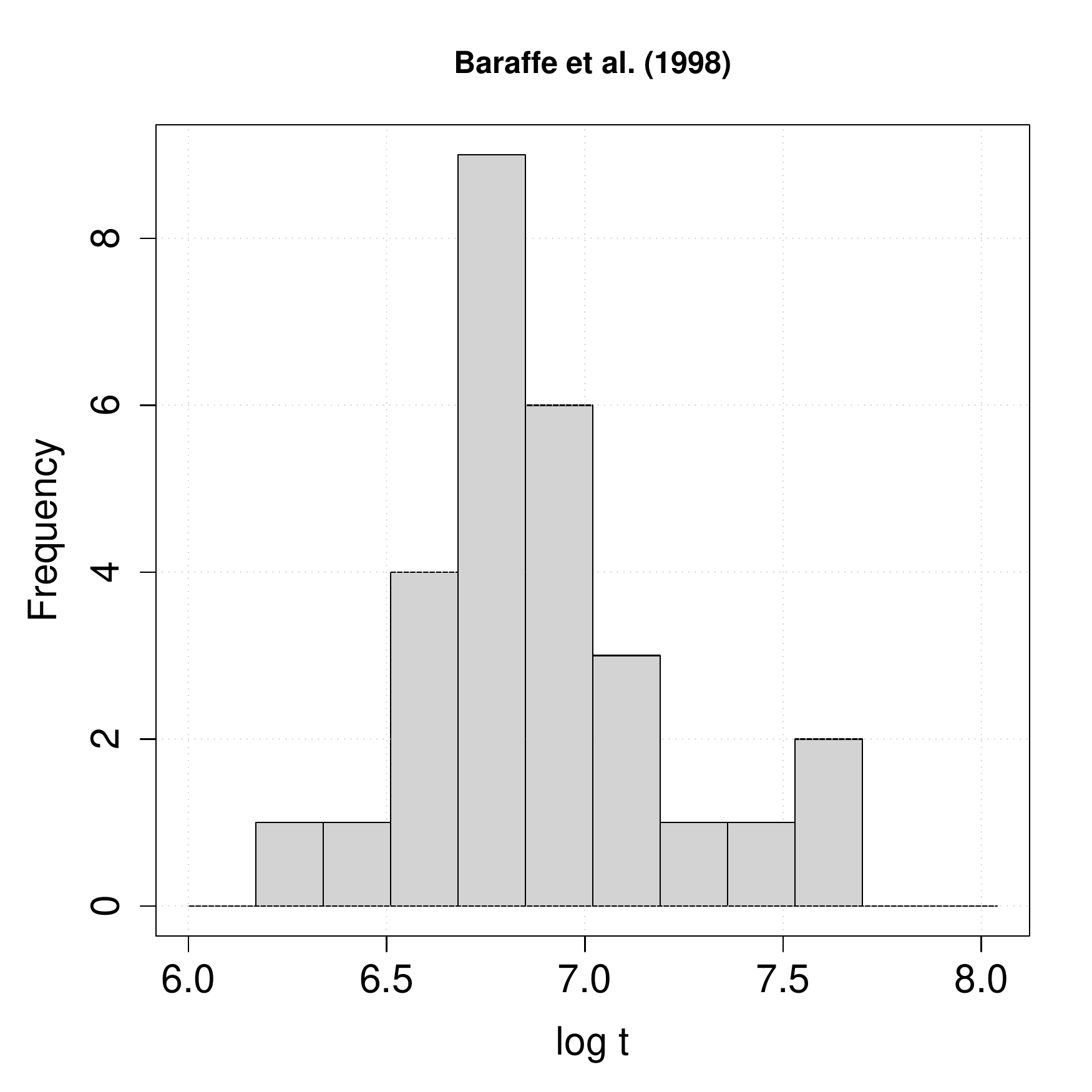}
\includegraphics[width=0.45\textwidth]{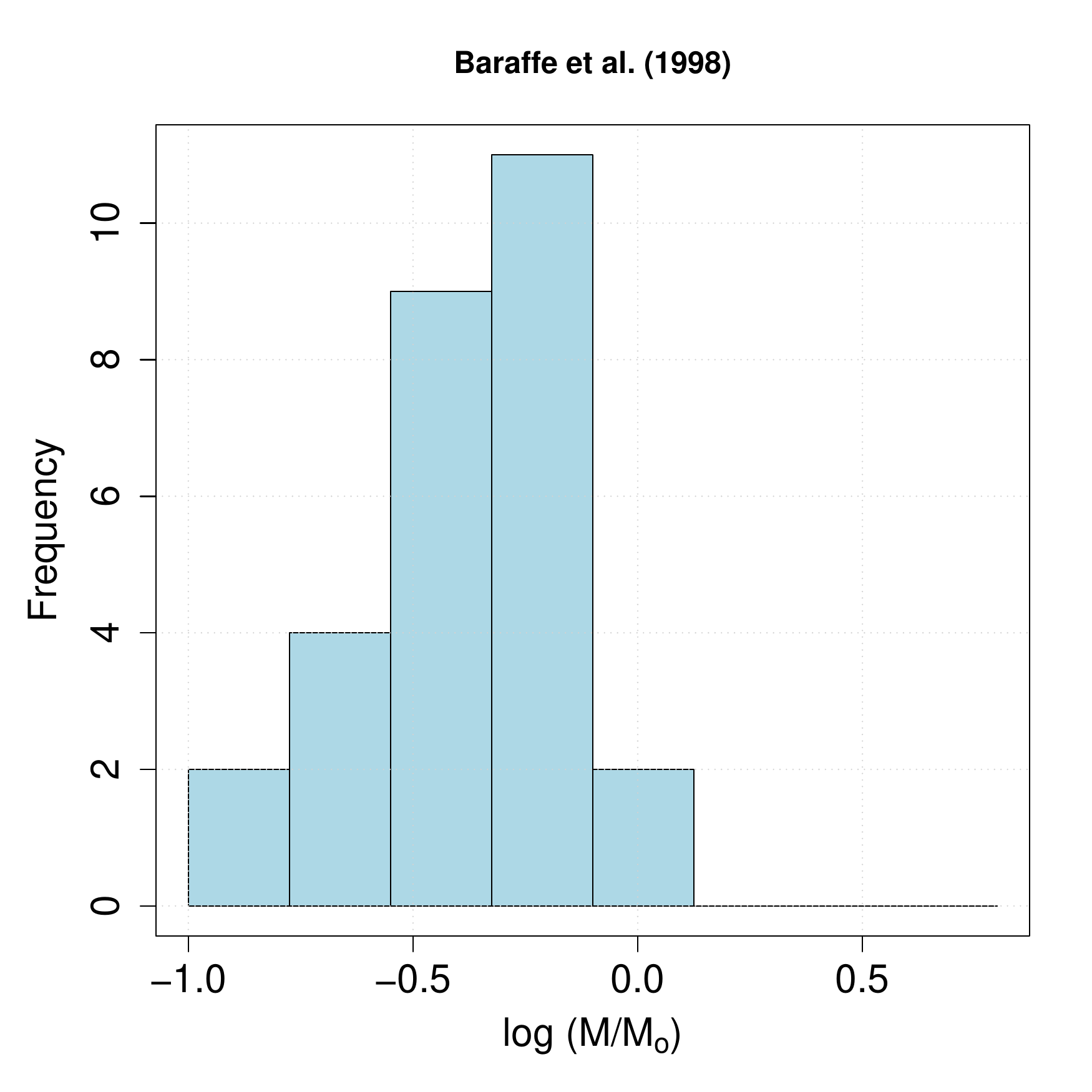}

\caption{Age (in $\log t$) and mass (in $\log M$) distributions for the TWA stars listed in Table~5 obtained with the grid of evolutionary models computed by \citealt{Siess(2000)} (\textit{upper panels}) and \citealt{Baraffe(1998)} (\textit{lower panels}). TWA~9A, TWA~21 and TWA~23 are the stars plotted near $\log t=7.5$, and TWA~11A is the star plotted near $\log(M/M_{\odot})$ = 0.5 in the upper panels. The Barrafe tracks do not cover such large masses.}
\label{fig_age_mass}
\end{center}
\end{figure*}

The HR-diagram presented in Figure~\ref{HRD} shows that most TWA members lie between the 3~Myr and 10~Myr isochrones. We derive the masses and ages of individual stars based on their position in the HR-diagram. The models available for PMS stars below 20 Myr are rather uncertain \citep{Soderblom(2013)} so we decided to test two of the more commonly used evolutionary models. To do so, we use the grid of pre-main sequence evolutionary tracks and isochrones computed by \citet{Siess(2000)} and \citet{Baraffe(1998)}. We use the version of the Baraffe et al. models with helium abundance $Y=0.282$ and mixing length parameter of 1.9, because of its success in confirming coevality of the components in young multiple systems as reported by \citet{White(1999)}. The stellar parameters that result from this investigation are presented in Table~\ref{mass_age}.We note that the stellar ages of the two components of the binary stars listed in Table~\ref{mass_age} agree with each other within their errors. The age and mass distributions of TWA moving group members are shown in Fig.~\ref{fig_age_mass}. 

We note from Fig.~\ref{fig_age_mass} and Table~\ref{mass_age} that TWA~9A, TWA~21 and TWA~23 are older than other moving group members according to both models. 
Concerning TWA~9A, we noticed that its kinematical distance calculated from the spatial velocity of the association as described in section 4.3 ($\pi_{kin} = 14.2 \pm 1.0$ mas) and its trigonometric distance ($\pi = 21.4 \pm 2.5$ mas) are very different. \citet{Weinberger(2013)} consider TWA 9A in their Section 4.2 and reject it as a member. Therefore we strongly suspect TWA~9A not to be TWA member or its various trigonometric parallax measurements (this work, Hipparcos) to be false. In any case this star has been rejected from the trace-back and is eliminated at a 3 $\sigma$ rejection test when calculating the mean evolutionary age of TWA. Removing TWA~9A, TWA~21 and TWA~23 from our CP analysis in Sect.~\ref{cp} has negligible effect on the CP position, thus they can still be regarded as kinematic members of the TWA. We also note from Table~\ref{mass_age}  that TWA 11A stands out on the mass distribution being more massive than other TWA members (see Fig.~\ref{fig_age_mass}). This star does not appear in Fig.~\ref{HRD} because of the chosen range of temperature and luminosity. 

The mean age for TWA stars derived in this work from the \citet{Baraffe(1998)} isochrones ($8.2\pm0.7$~Myr) is consistent with the mean age of 9~Myr derived by \citet{Weinberger(2013)} in their analysis. However, the median age of $10.1$~Myr reported in that analysis differs from the value of $7.2$~Myr obtained in this work. The difference between these two age estimates can be due to some combination of a different sample of TWA stars, the effect of stellar extinction (not considered by \citealt{Weinberger(2013)}) and different ways of converting observed data (magnitudes, colors and spectral types) to theoretical values (effective temperature and luminosity). We present in Table~6 the mean and median age derived for each set of isochrones used in this work after a $3\sigma$ elimination.

We note from Table~6  that the age estimates inferred from \citet{Siess(2000)} isochrones are systematically smaller than the results given by \citet{Baraffe(1998)}. That the mean and median age results listed in Table~6 for each set of isochrones differ, is not surprising given the different input assumptions and stellar physics in each model.  

The mean of $8.2\pm0.7$~Myr inferred from the \citet{Baraffe(1998)} isochrones (after $3\sigma$ elimination) is compatible with the dynamical traceback age within $1\sigma$ of the computed errors. Furthermore, the mean age of $7.3\pm1.0$~Myr calculated using only the traceback-core stars included in Table~5 is also fully consistent with the dynamical age and provides a more direct comparison of these age estimates. These results agree well with previous estimates using different strategies as summarized in Table~2 of \citet{Fernandez(2008)}. We conclude from this analysis that the dynamical traceback age derived in Sect.~5 is compatible with the isochronal age estimates derived from both evolutionary models used in this work, but it is in better agreement with the results inferred from the \citet{Baraffe(1998)} isochrones.

%FIGURE 6

%FIGURE 7

%TABLE 6
\begin{table*}[!htp]
\begin{center}
\caption{\label{age_results}
Comparison of age results inferred from the evolutionary models of \citet{Siess(2000)} and \citet{Baraffe(1998)}.}
\resizebox{12cm}{!} {
\begin{tabular}{lcccccccc}
\hline
&\multicolumn{3}{c}{All stars}&\multicolumn{3}{c}{Traceback core stars}\\
\hline
&Nb Stars&Mean age&Median age&Nb Stars&Mean age&Medianage\\
&              &\multicolumn{2}{c}{[Myr]}&           &\multicolumn{2}{c}{[Myr]}\\
\hline
\citet{Siess(2000)}   &15&$5.8\pm0.3$&5.4&11&$5.9\pm0.4$&6.0\\
\citet{Baraffe(1998)} &17&$8.2\pm0.7$&7.2&13&$7.3\pm1.0$&6.3\\
\hline
\end{tabular}
}
\end{center}
\tablefoot{We provide the number of stars, mean age and median age derived from the evolutionary models of \citet{Siess(2000)} and \citet{Baraffe(1998)} after a 3 $\sigma$ elimination for (i) all members of the association (marked with `Y' and `Y?' in Table~3), and (ii)  the traceback-core stars listed in Sect.~5.2. 
\vspace{1cm}
}
\end{table*}

%----------------------------------------------------------------------------------------------------------------
%						CONCLUSIONS
%----------------------------------------------------------------------------------------------------------------
\section{Conclusions}

Based on astrometric and photometric observations performed with the ESO/NTT telescope we present  trigonometric parallaxes, proper motions and photometry for 13 stars in the TW Hydrae Association (TWA). This represents a gain of almost a factor of 3 when compared to the situation in the \textit{Hipparcos} era (only five stars with trigonometric parallaxes). The average precision of the parallaxes derived in the present study is 2.2~mas. Our proper motions are in good agreement with those given in UCAC4 and SPM4. 

Using information provided in current astrometric catalogs and recent papers we set up an updated database for 34 previously proposed TWA members that allowed us to completely revisit the kinematics of the association. 

Based on a convergent point search method we identify a moving group with 31 members and derive kinematic parallaxes for 7 of these group members with unknown trigonometric parallaxes. 

We derive the space motion for individual stars of the moving group and for 16 members trace motions back in time to seek the time these stars occupied a minimum volume. We derive a dynamical age of $7.5\pm0.7$~Myr for the association.

Using the parallaxes derived in this paper and published data we estimate mass and age for TWA moving group members from pre-main sequence evolutionary models and find a mean age  for the members as defined in Sect.~4.4 of $5.8\pm0.3$~Myr (\citet{Siess(2000)}) and of $8.2\pm0.7$~Myr (\citet{Baraffe(1998)}). 

To conclude, we show that the dynamical age of the association obtained via the traceback technique and the average ages derived from theoretical evolutionary models are compatible and that the \citet{Baraffe(1998)} mean age is in excellent agreement (within 1 $\sigma$). We observe that the theoretical ages derived are dependent on multiple parameters and differ when applied to the same observational data while the dynamical age relies only on 	astrometric data and appears eventually more reliable.

%----------------------------------------------------------------------------------------------------------------
%							ACKNOWLEDGMENTS
%----------------------------------------------------------------------------------------------------------------
\begin{acknowledgements} 
We thank the referee for constructive comments that improved the paper. We also acknowledge
partial financial support from the Brazilian organization FAPESP and CAPES, and the French organization COFECUB.
\end{acknowledgements}

%----------------------------------------------------------------------------------------------------------------
%							BIBLIOGRAPHY
%----------------------------------------------------------------------------------------------------------------
\bibliographystyle{aa}
\bibliography{references.bib}

\end{document}